\documentclass[article]{new-aiaa}
\usepackage[utf8]{inputenc}
\usepackage{textcomp}
\usepackage{xcolor}
\usepackage{graphicx}
\usepackage{amsmath}
\usepackage[version=4]{mhchem}
\usepackage{siunitx}
\usepackage{longtable,tabularx}
\usepackage{appendix}
\usepackage{multirow}
\usepackage{float}
\usepackage{tikz}
\usepackage{algorithm}
\usepackage{algpseudocode}

\usepackage{sectsty}
\usepackage{indentfirst}
\renewcommand\thesection{\arabic{section}}
\renewcommand\thesubsection{\thesection.\arabic{subsection}}
\usepackage{setspace}
\sectionfont{\fontsize{14}{12}\selectfont}
\subsectionfont{\fontsize{13}{12}\selectfont}
\subsubsectionfont{\fontsize{12}{12}\selectfont}

\providecommand{\keywords}[1]
{
  \small	
  \textbf{\textit{Keywords---}} #1
}

\newcommand*\circled[1]{\tikz[baseline=(char.base)]{
            \node[shape=circle,draw,inner sep=1pt] (char) {#1};}}

\title{A Multi-physics Model of Flow from Coronary Angiography: Insights to Microvascular Function}
\author{\fontsize{10pt}{14pt}\selectfont Haizhou Yang\textsuperscript{1}, Jiyang Zhang\textsuperscript{1,2},  Ismael Z. Assi\textsuperscript{3}, Brahmajee K. Nallamothu\textsuperscript{4}, Krishna Garikipati\textsuperscript{5} and C. Alberto Figueroa\textsuperscript{1,6}\footnote{Professor, University of Michigan, figueroc@med.umich.edu}}

\affil{\fontsize{10pt}{14pt}\selectfont \textsuperscript{1}Department of Biomedical Engineering, University of Michigan, Ann Arbor, MI 48109, USA}
\affil{\fontsize{10pt}{14pt}\selectfont \textsuperscript{2}Department of Mechanical Science and Engineering, Sichuan University, Chengdu, Sichuan 610017, China}
\affil{\fontsize{10pt}{14pt}\selectfont \textsuperscript{3}College of Medicine, University of Cincinnati, Cincinnati, OH 45221, USA}
\affil{\fontsize{10pt}{14pt}\selectfont \textsuperscript{4}Department of Internal Medicine, University of Michigan, Ann Arbor, MI 48109, USA}
\affil{\fontsize{10pt}{14pt}\selectfont \textsuperscript{5}Department of Aerospace and Mechanical Engineering, University of Southern California, Los Angeles, CA 90089, USA}
\affil{\fontsize{10pt}{14pt}\selectfont \textsuperscript{6}Department of Surgery, University of Michigan, Ann Arbor, MI 48109, USA}

\begin{document}

\maketitle

\begin{abstract}

Coronary Microvascular Dysfunction (CMD) is characterized by impaired vasodilation and can lead to insufficient blood flow to the myocardium during stress or exertion, affecting millions of people globally. Despite their diagnostic value, invasive, wire-based diagnosis techniques of CMD, such as index of microcirculatory resistance (IMR) and coronary flow reserve (CFR), are underutilized due to their complexity and inconsistency. Coronary angiography, one of the most commonly used imaging modalities, offers valuable flow information that assists in diagnosing CMD. However, this information is not fully understood or utilized in current clinical practice. In this study, a 3D-0D coupled multi-physics computational fluid dynamics (CFD) model was developed and calibrated to simulate and study the process of contrast injection and washout during clinical angiography. A contrast intensity profile (CIP) was introduced to describe the dynamics of coronary angiography data. Additionally, sensitivity studies were conducted to evaluate the influence of various coronary lumped parameter model (LPM) parameters on the shapes of CIPs. The results demonstrate that the multi-physics model can be effectively calibrated to produce physiologically meaningful hemodynamic results. Sensitivity studies reveal that resistance has a greater impact on the rising and falling slopes of CIP than capacitance, with higher resistance amplifying this effect. The model and results are presented here. These results are potentially transformative, as they provide a tool for interpreting angiographic data and ultimately extracting information concerning coronary microcirculation.

\end{abstract}


\section{Introduction}

Coronary artery disease (CAD) and coronary microvascular dysfunction (CMD) both alter blood supply to the heart but involve different mechanisms and diagnostic challenges. CAD affects the large epicardial vessels, where atherosclerotic plaques (stenoses) may reduce blood flow and oxygen delivery to the myocardium, leading to angina or heart attacks. Anatomic diagnosis of CAD with coronary angiography in the cardiac catheterization lab (“cath lab”) has been the cornerstone of its evaluation for decades. In recent years, however, functional diagnosis has transformed its assessment via invasive, wire-based tools (e.g., fractional flow reserve(FFR)) and newer, non-wire-based computational methods (e.g., AngioFFR) \cite{fearon2018clinical,morris2020angiography,davies2017use}. In contrast, CMD impacts the microvasculature and primarily leads to endothelial dysfunction, inflammation, and hormonal imbalances that affect vasodilatory capacity with insufficient blood flow during stress or exertion. CMD is a more common problem in women \cite{bairey1999women,sucato2022classification,safdar2020prevalence} and frequently presents without CAD. Anatomic diagnosis of CMD is currently not possible, while its functional diagnosis is a challenging task in routine clinical settings \cite{camici2007coronary,godo2021coronary,sucato2022classification}.

Typically, the diagnosis of CMD requires advanced perfusion imaging outside of the cath lab to uncover myocardial territories with insufficient flow along with symptoms through tools such as cardiac positron emission tomography (PET) or magnetic resonance imaging (MRI) \cite{sucato2022classification, lanza2010primary,el2020myocardial,maron2020initial}. Invasive, wire-based techniques for assessing CMD in the cath lab, such as the index of microcirculatory resistance (IMR) \cite{martinez2015index,ng2012index}, have recently become available and are receiving growing attention. These techniques utilize specialized equipment to simultaneously measure pressure and estimate flow after placing a wire in the coronary vessel \cite{aarnoudse2004epicardial,fearon2013prognostic,lee2015invasive,taqueti2018coronary}. Although they enable a complete physiologic assessment of CAD and CMD by measuring FFR, IMR, and coronary flow reserve (CFR), the measurement of IMR and CFR via wire-based techniques are infrequently utilized due to their complexity and inconsistency \cite{fearon2013prognostic,lee2015integrated,desai2015appropriate}. Given the challenges in diagnosing CMD overall and independently from CAD, there is a pressing need to develop computational methods for CMD assessment, particularly in the context of coronary angiography performed in the cath lab.

\subsection{Coronary Angiography}
Coronary angiography systems are essential tools widely available for diagnostic and interventional procedures in cardiology. Coronary angiography is considered the gold standard for coronary imaging, providing high-resolution images (0.1 mm) at a high frame rate of 10-15 frames per second \cite{abdelaal2014effectiveness}. These systems visualize the coronary arteries using X-rays combined with a radio-opaque contrast agent (dye) that highlights blood flow. The contrast agent is injected through the coronary artery ostia, either on the left or right side. After the contrast injection stops, it washes through the coronary tree over several seconds. Injections are performed under two hemodynamic conditions – rest and hyperemia. Typically, 4 to 5 angiogram videos from different viewpoints are captured for the left coronary tree and 2 to 3 for the right coronary tree. 

Despite its high spatial and temporal resolutions and widespread utilization, coronary angiography has a long-recognized low diagnostic yield and currently provides little functional data on CMD \cite{patel2010low}. Information is often limited to the characterization of epicardial lesions (\% stenosis) and subjective, qualitative assessment of flow via TIMI (Thrombolysis in Myocardial Infarction) scores in the setting of acute coronary syndromes\cite{doherty2021predictors}. One such score is the \textit{TIMI flow grade}, which provides a subjective grading system to evaluate the level of perfusion through a blocked or narrowed artery. The grading scale ranges from 0 to 3. A grade of 0 indicates a lack of flow beyond a certain point. On the other hand, a grade 3 indicates normal flow and complete perfusion through the coronary artery. A slightly more quantitative score of coronary flow is the \textit{TIMI frame count} \cite{gibson1996timi,gibson2004coronary}. Here, an analyst counts the number of frames it takes the contrast agent to reach a selected distal landmark. A higher speed of blood corresponds to a lower frame count, and vice versa. This method provides somewhat quantitative information, but does not directly measure flow and requires manual inputs from the operator. Fundamentally, the analysis of angiographic data has seen limited evolution in the last decades. We therefore submit that there is a pressing need to develop formulations to extract quantitative (rather than qualitative) information encoded in the dynamic coronary angiography data.

\subsection{Contrast Intensity Profiles derived from Coronary Angiography}

\begin{figure}[h]
\centering
\includegraphics[width=0.7\textwidth]{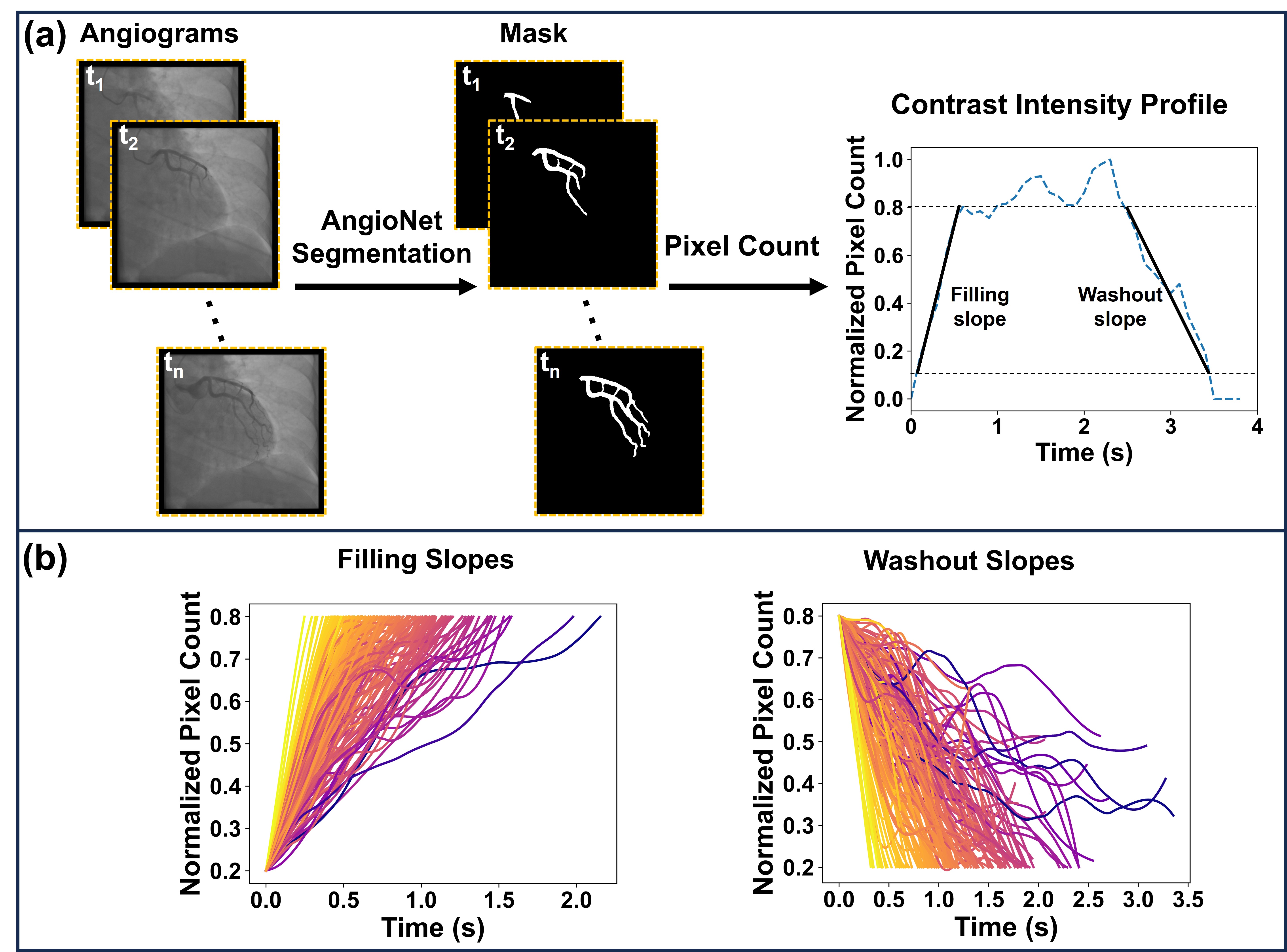}
\caption{\centering (a) Contrast intensity profile (CIP) generation from clinical angiography data in a cohort of patients without epicardial disease referred for diagnostic coronary angiography. (b) Filling and washout slopes of the CIPs for a subset of the patient cohort.}
\label{fig:Jesse_paper}
\end{figure}

Our group has recently developed AngioNet \cite{iyer2021angionet,stevens2021}, a convolutional neural network for vessel segmentation of coronary angiography images. More recently, we have applied AngioNet to study the dynamics of coronary flow in a cohort of patients without epicardial disease referred for a diagnostic coronary angiography \cite{resnick2024neural}. Fig. \ref{fig:Jesse_paper}(a) describes the analysis process. A time-series of $n$ angiograms is processed using AngioNet to segment the portions of the vessels filled with iodine contrast and create segmentation masks. Then, the number of bright pixels in the segmented images is counted for each frame, producing a contrast intensity profile (CIP) that reflects the injection and washout of the contrast agent. A CIP curve shows three distinct stages: a filling stage, a plateau corresponding to the frames in which the coronary tree is fully saturated with contrast, and a washout stage, with the filling (raising) and washout (falling) slopes (black solid lines) of the CIP.  Fig. \ref{fig:Jesse_paper}(b) shows filling and washout slopes for a subset of the patient cohort. Warmer colors indicate faster filling and emptying dynamics of contrast.  Our results showed a strong correlation between filling slopes and \textit{TIMI frame count} measured by two independent experts \cite{resnick2024neural}. A large variability in the washout slopes is observed. 

Numerous studies have demonstrated the strong link between features in physiological measurements and physical properties such as pulse wave analysis of aortic pressure to estimate compliance, resistance, forward and backward traveling waves, etc \cite{Denardo2010,Westerhof1972,Nichols2008}. In this paper, we hypothesize that the washout slopes of the CIPs are determined by the properties of the microcirculation: higher microvascular resistance would result in slower washout slopes, whereas lower microvascular resistance would lead to faster washout slopes. 

To test this hypothesis, in this paper we develop and present a computational multi-physics model of contrast injection to study the relationship between CIP features and microvascular parameters. We will demonstrate that: 1) The multi-physics model can be effectively calibrated to match arbitrary sets of clinical data consisting of coronary angiograms and data on flow and pressure, under resting and hyperemic conditions. The model produces computational CIPs which closely mimic CIPs derived from clinical coronary angiography data (see Fig. \ref{fig:Jesse_paper}). 2) Coronary microvascular resistances of the multi-physics model determine the slopes of the CIPs. We submit that this model is potentially transformative, as it provides a tool for interpreting angiographic data and ultimately extracting information concerning coronary microcirculation. Furthermore, this paper is, to our knowledge, the first effort to simulate and analyze the complex physics of coronary angiography with a high-fidelity computational model.

The rest of this paper is structured as follows. Section \ref{Methodology} introduces a multi-physics CFD model for simulating coronary angiography and describes the process to perform model calibration. Sections \ref{Results} and \ref{Discussion} present the results and discussions, respectively. The paper concludes with a summary of the findings and provides directions for future research in Section \ref{Conclusion}.

\section{Methodology}\label{Methodology}

\subsection{Patient Data}
The clinical data used in this study corresponds to a 64-year-old female with suspected CMD. Imaging data consisted of coronary angiograms under resting and hyperemic conditions, as well as coronary computed tomography angiography (CCTA). Right and left coronary angiography was performed in multiple views by hand injections of Isovue 370. In this study, we limited our analysis to the left coronary angiography. Hyperemia was induced through 140 $\mu$g/kg/min of adenosine. A coronary flow reserve (CFR), defined as the ratio of hyperemia to resting flow rates, of 2.2  was measured in the left anterior descending (LAD). Heart rates of 60 bpm (rest) and 82 bpm (hyperemia) were extracted from the corresponding angiograms. The CCTA data included a volume of $320 \text{mm} \times 320 \text{mm} \times 268.5 \text{mm}$, with an in-plane resolution of  $0.625 \text{mm} \times 0.625 \text{mm}$ and a through-plane resolution of 1.5 mm. Angiograms taken under resting conditions were composed of $512 \times 512$ images with an in-plane resolution of $0.368 \text{mm} \times 0.368 \text{mm}$ and a frame rate of 10 fps. For angiograms under hyperemic conditions, the images were $512 \times 512$ with an in-plane resolution of $0.279 \text{mm} \times 0.279 \text{mm}$ and a frame rate of 7.5 fps. This study was approved by the University of Michigan institutional review board (IRB-HUM00155491).

\subsection{Multi-physics CFD Model of Contrast Injection}

\begin{figure}[h]
\centering
\includegraphics[width=0.8\textwidth]{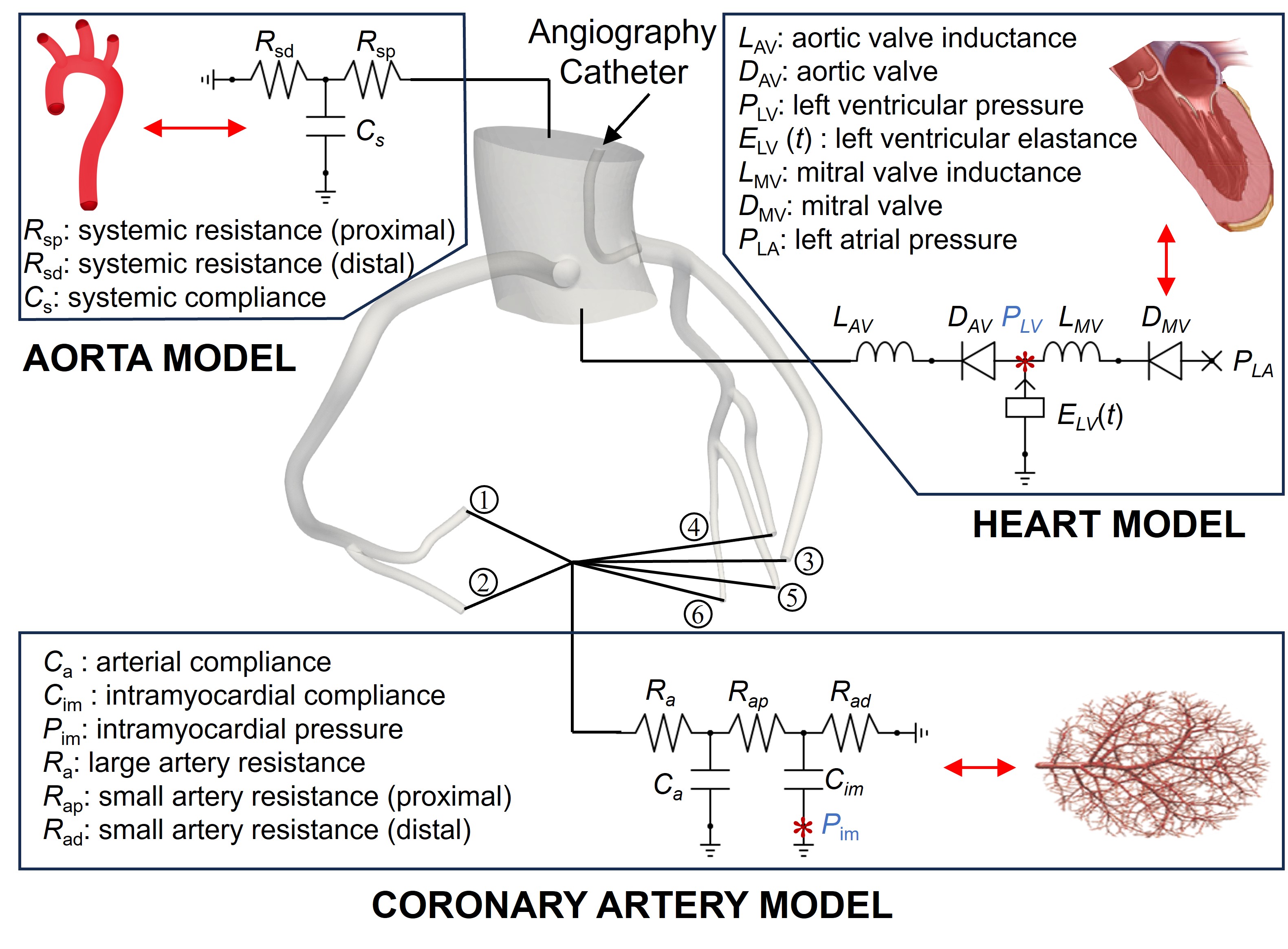}
\caption{\centering 3D-0D Multi-physics model of contrast injection.}
\label{fig:CFD_model}
\end{figure}

We developed a novel multi-physics CFD model of iodine contrast injection using CRIMSON, an open-source software for patient-specific hemodynamic simulation \cite{arthurs2021crimson}. The model simultaneously solves for hemodynamics (e.g., blood flow) and advection-diffusion transport of a scalar quantity (e.g., the contrast agent). The hemodynamics model relies on the widely established 3D-0D modeling approach to represent blood flow in the larger coronary arteries (3D component of the model), as well as the coronary microcirculation (0D or lumped parameter model (LPM) component of the model). Fig. \ref{fig:CFD_model} depicts this multi-physics model.  An anatomical model of the aortic root along with the main branches of both the left and right coronary arteries was constructed from the CCTA data. The right coronary tree consists of the right coronary artery (\circled{1} RCA) and the acute marginal (\circled{2} AM) branches. The left coronary tree consists of four branches: the left anterior descending (\circled{3} LAD), obtuse marginal (\circled{4} OM1, \circled{5} OM2), and left circumflex (\circled{6} LCx) arteries. This model also includes the distal end of an angiography catheter, positioned facing the ostia of the left coronary artery, which is used to simulate the injection of the contrast agent. There are three different 0D (lumped-parameter) models: 1) an Aorta model, represented by a 3-element Windkessel, coupled to the outlet face of the aorta; 2) a Heart Model, represented by a circuit describing the dynamics of ventricular pumping, coupled to the inlet face of the aorta; and 3) a series of Coronary Models coupled to each of the outlet faces of the six coronary branches above. 

\begin{figure}[H]
\centering
\includegraphics[width=0.4\textwidth]{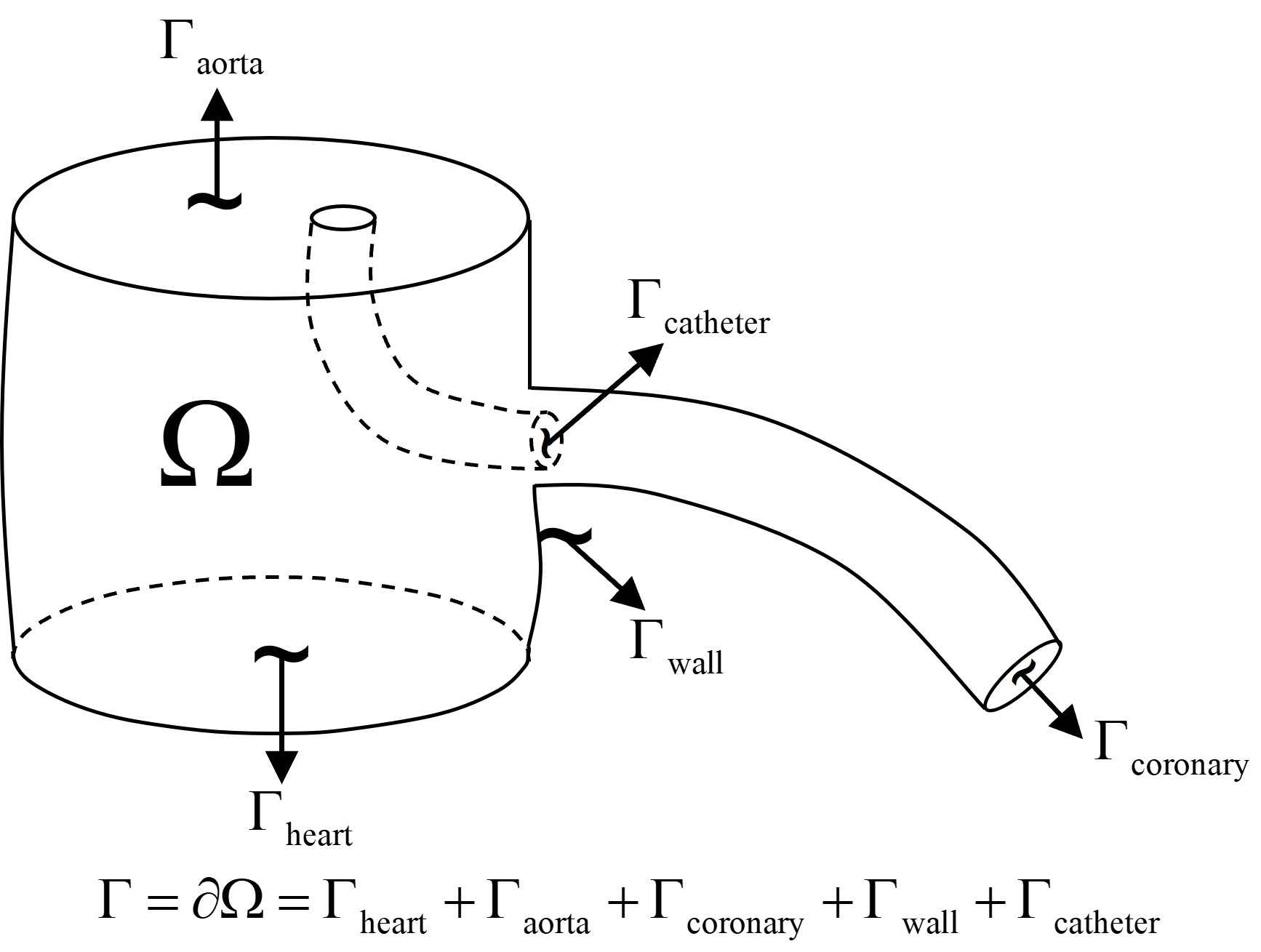}
\caption{\centering Schematic of computational domain $\Omega$ and it boundary $\Gamma$.}
\label{fig:bcs}
\end{figure}

\underline{Governing equations:} This model employs the finite element method to solve the underlying incompressible Navier–Stokes and advection-diffusion transport equations within the computational domain ($\Omega$), calculating velocity \textbf{u}, pressure $p$, and iodine contrast concentration $c$:

\begin{equation}
\label{eqn:NS}
\begin{aligned}
&   \left\{
\begin{aligned}
&\nabla \cdot \mathbf{u} = 0 \\
&\frac{\partial \mathbf{u}}{\partial t} + (\mathbf{u} \cdot \nabla) \mathbf{u} = -\frac{1}{\rho} \nabla p + \nu \nabla^2 \mathbf{u}
\end{aligned}
\right. \quad\text{on}\quad \Omega,\\
\end{aligned}
\end{equation}

\begin{equation}
\label{eqn:CD}
\frac{\partial c}{\partial t} + \nabla \cdot (\mathbf{u} c) = D \nabla^2 c\quad\text{on}\quad \Omega,
\end{equation}
where $\rho$ denotes the fluid density, $\nu$ the kinematic viscosity, and $D$ the diffusion coefficient. The boundary $\Gamma$ of the computational domain $\Omega$ is $\Gamma=\partial\Omega=\Gamma_\text{heart}+\Gamma_\text{aorta}+ \Gamma_\text{coronary}+\Gamma_\text{catheter}+\Gamma_\text{wall}$, see Fig. \ref{fig:bcs}.  $\Gamma_\text{heart}$ corresponds to the aortic inlet,  $\Gamma_\text{aorta}$ and $\Gamma_\text{coronary}$ correspond to the aortic outlet and coronary outlet boundaries, respectively. $\Gamma_\text{catheter}$ represents the face through which contrast is injected, and $\Gamma_\text{wall}$ corresponds to the vessel wall boundaries. 

The fluid boundary conditions for the Navier–Stokes equation are
\begin{equation}
\label{eqn:fluid}
\begin{aligned}
&   \left\{
\begin{aligned}
& \text{Lumped Parameter Heart Model} \quad\text{on}\quad \Gamma_\text{heart}\\ %p(\textbf{x},t) = \text{LPM}(\textbf{x},t) \quad \text{or} \quad \textbf{u}(\textbf{x},t) = 0
& \text{Lumped Parameter Aorta Model} \quad\text{on}\quad \Gamma_\text{aorta}\\
& \text{Lumped Parameter Coronary Artery Model} \quad\text{on}\quad  \Gamma_\text{coronary}\\
& \int_{\Gamma_\text{catheter}}{\mathbf{u}} dS = Q_{\text{catheter}}(t) \quad \text{on} \quad  \Gamma_\text{catheter}\\
& \textbf{u}(\textbf{x},t) = 0 \quad\text{on}\quad \Gamma_\text{wall}
\end{aligned}
\right.
\end{aligned}
\end{equation}
for the sake of notational simplicity. The reader is referred to references \cite{arthurs2021crimson,pope2008estimation,kim2010patient,algranati2010mechanisms,sankaran2012patient,westerhof2009arterial} for details on the formal formulation of these models.
The  scalar boundary conditions for the advection-diffusion equation are
\begin{equation}
\label{eqn:scalar}
\begin{aligned}
&   \left\{
\begin{aligned}
& c(\textbf{x},t) = 0 \quad\text{on}\quad \Gamma_\text{heart}\\
& \textbf{n}\cdot(-D \nabla c) = 0 \quad\text{on}\quad \Gamma_\text{aorta}\cup\Gamma_\text{coronary}\\
& c(\textbf{x},t) = c_\text{0} \quad \text{on} \quad  \Gamma_\text{catheter}\\
& \textbf{n}\cdot(-D \nabla c+\mathbf{u} c) = 0 \quad\text{on}\quad \Gamma_\text{wall}
\end{aligned}
\right.
\end{aligned}
\end{equation}
Furthermore, Neumann boundary condition stabilization at the outlets was implemented based on the methods described in \cite{hughes2005conservation,hsu2010improving,lynch2020numerical}.

\begin{figure}[H]
\centering
\includegraphics[width=0.4\textwidth]{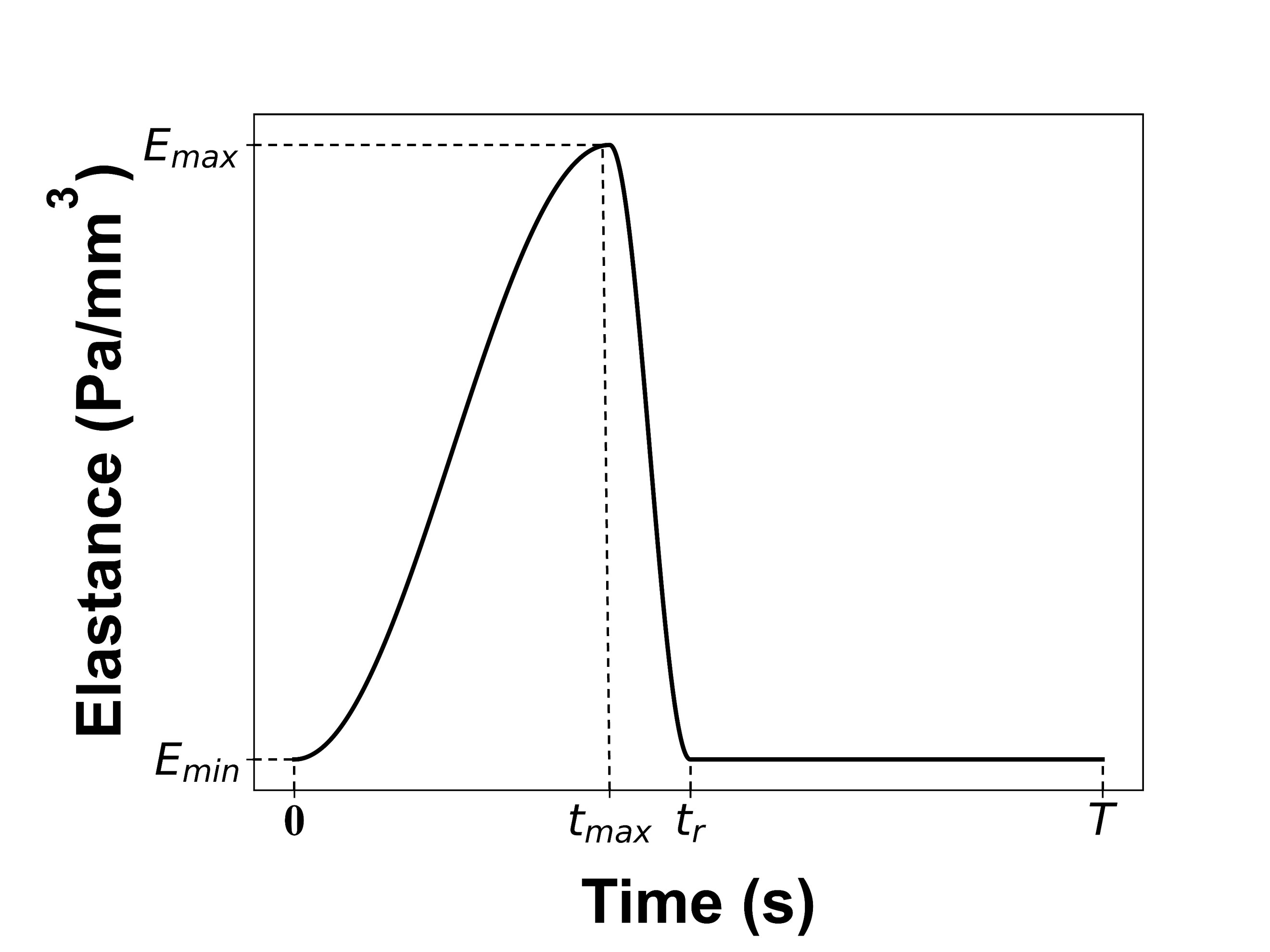}
\caption{\centering Elastance function.}
\label{fig:elastance}
\end{figure}

\underline{Heart LPM:} A key aspect of the model is the left heart LPM \cite{arthurs2016mathematical} coupled to the aortic inlet $\Gamma_\text{heart}$. The Heart LPM consists of: an aortic valve inductance $L_\text{AV}$, an aortic valve $D_\text{AV}$, a left ventricular elastance function $E_\text{LV}(t)$,  a mitral valve inductance $L_\text{MV}$, a mitral valve $D_\text{MV}$, and a fixed left atrial pressure $P_\text{LA}$. The time-varying elastance function $E_\text{LV}(t)$ represents ventricular contractility \cite{pope2008estimation}, as shown in Fig. \ref{fig:elastance}. Mathematically, $E_\text{LV}(t)$ is the relationship between the volume and the pressure of the left ventricle. The function is defined by five variables: maximum elastance $E_\text{max}$, minimum elastance $E_\text{min}$, cardiac cycle $T$, time to maximum elastance $t_\text{max}$, and relaxation time $t_\text{r}$. The time interval 0 to $t_\text{max}$ corresponds to the systolic phase, when the ventricle contracts and pumps blood into the systemic circulation, while the time interval $t_\text{max}$ to $T$ represents the diastolic phase, during which the ventricle relaxes and fills with blood. 

\underline{Coronary LPM:} Another important component of the model is the coronary LPMs coupled to each of the outlet faces defining $\Gamma_\text{coronary}$ \cite{kim2010patient}. This model accounts for myocardial compression, leading to the characteristic diastolic-dominated coronary flow waveforms. The model is defined by: a large artery resistance $R_\text{a}$, a proximal small artery resistance $R_\text{ap}$, a distal small artery resistance $R_\text{ad}$, an arterial compliance $C_\text{a}$, and an intramyocardial compliance $C_\text{im}$. An intramyocardial pressure $P_\text{im}$ is set at the node distal to the intramyocardial compliance $C_\text{im}$. The intramyocardial pressure is closely tied to ventricular pressure, as well as other mechanisms \cite{algranati2010mechanisms}, and can be expressed as $P_\text{im} = \alpha P_\text{LV}$, where $\alpha$ is a transfer coefficient \cite{sankaran2012patient}. As the left and right coronary arteries reside in different regions of the myocardium, they are exposed to different intramyocardial pressures. Here, $\alpha$ was set to 0.5 and 0.25 for the left and right coronary arteries, respectively.

\underline{Aorta LPM:} The outlet of the aortic root $\Gamma_\text{aorta}$ is connected to an aorta LPM,  represented by a three-element Windkessel model \cite{westerhof2009arterial}. This model consists of: a systemic compliance $C_\text{s}$, a proximal systemic resistance $R_\text{sp}$, and a distal systemic resistance $R_\text{sd}$.

\begin{figure}[h]
\centering
\includegraphics[width=0.4\textwidth]{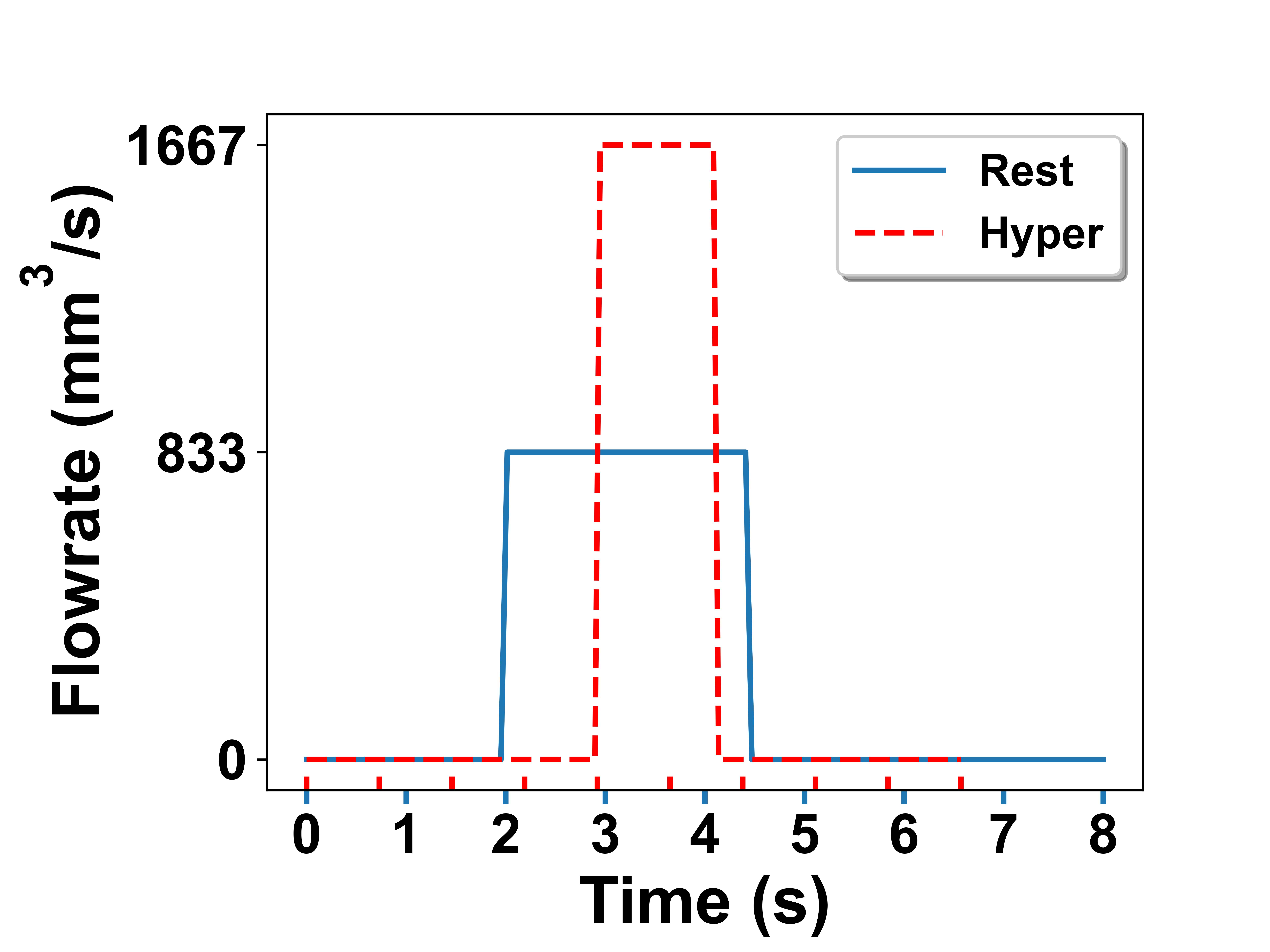}
\caption{\centering Profiles of contrast agent injection for rest and hyperemia.}
\label{fig:Catheter_flow}
\end{figure}

\underline{Contrast injection parameters:} A 2 mm rigid catheter inserted in the aorta, and facing the ostia of the left coronary coronary tree was included in the anatomical model. The position of the catheter was assumed fixed through the simulation. 
The function $ Q_{\text{catheter}}(t)$ defining contrast injection through the catheter was specified by: i) assuming that the total volume of injected contrast agent was the same for both rest and hyperemia, and ii) estimating the duration of contrast injection from the angiograms.
$Q_{\text{catheter}}(t)$ are therefore step functions, as illustrated in Fig. \ref{fig:Catheter_flow}. The total volume of contrast agent injected was 2 milliliters, representative of actual angiography procedures \cite{sacha2019ultra}. A contrast agent concentration of $c_{0} = 400$ $\text{mg/ml}$ with a diffusivity $D = 0.00203$ $\text{mm}^2/\text{s}$ \cite{caschera2016contrast} were assumed in both states. Under resting conditions, a constant flow rate of 833 $\text{mm}^3/\text{s}$ was injected over $2.4$ s. Under hyperemic conditions, a constant flow rate of 1667 $\text{mm}^3/\text{s}$ was injected over $1.2$ s.
% The contrast agent is injected at a constant rate $Q_\text{inj}$=833 under resting condition, and $Q_\text{inj}$=1667 under hyperemic condition, at a concentration $c_{0} = 400$ $\text{mg/ml}^3$, with a diffusivity $D = 0.00203$ $\text{mm}^2/\text{s}$ \cite{caschera2016contrast}, and over the time interval $t_\text{e}$ - $t_\text{s}$, where $t_\text{e}$ and $t_\text{s}$ represent the starting and ending times of injection, respectively. 

\underline{Simulation setup:} Blood was assumed to be Newtonian, with viscosity $\mu$ = 0.004 Pa·s and density $\rho$ = 1060 $\mathrm{kg/m}^3$. Vessel walls were assumed to be rigid. The estimated heart rates of 60 bpm and 82 bpm correspond to cardiac cycles of 1 s and 0.73 s for rest and hyperemia, respectively.  For the resting state, simulations were conducted over 8 cardiac cycles. The first two cycles were run to allow the hemodynamics to achieve cycle-to-cycle periodicity. Contrast agent injection started at the third cardiac cycle. For the hyperemic state, simulations were conducted over 9 cardiac cycles, spanning 6.57 s. Four cardiac cycles were first run to allow hemodynamics to achieve cycle-to-cycle periodicity. Contrast agent injection started at the fifth cardiac cycle. Timing of rest and hyperemia cardiac cycles is included in Fig. \ref{fig:Catheter_flow}. A time step size of 0.0001 s was used for simulating both states. A convergence tolerance of $1\text{e}^{-3}$ was set. Using anisotropic and curvature-based mesh refinement, and after performing a mesh-independence analysis, the computational domain was discretized into approximately 1.5M linear tetrahedral elements. Simulations were run for approximately 50 hours with the CRIMSON solver\cite{arthurs2021crimson} using 108 computing cores.

\begin{figure}[h]
\centering
\includegraphics[width=0.9\textwidth]{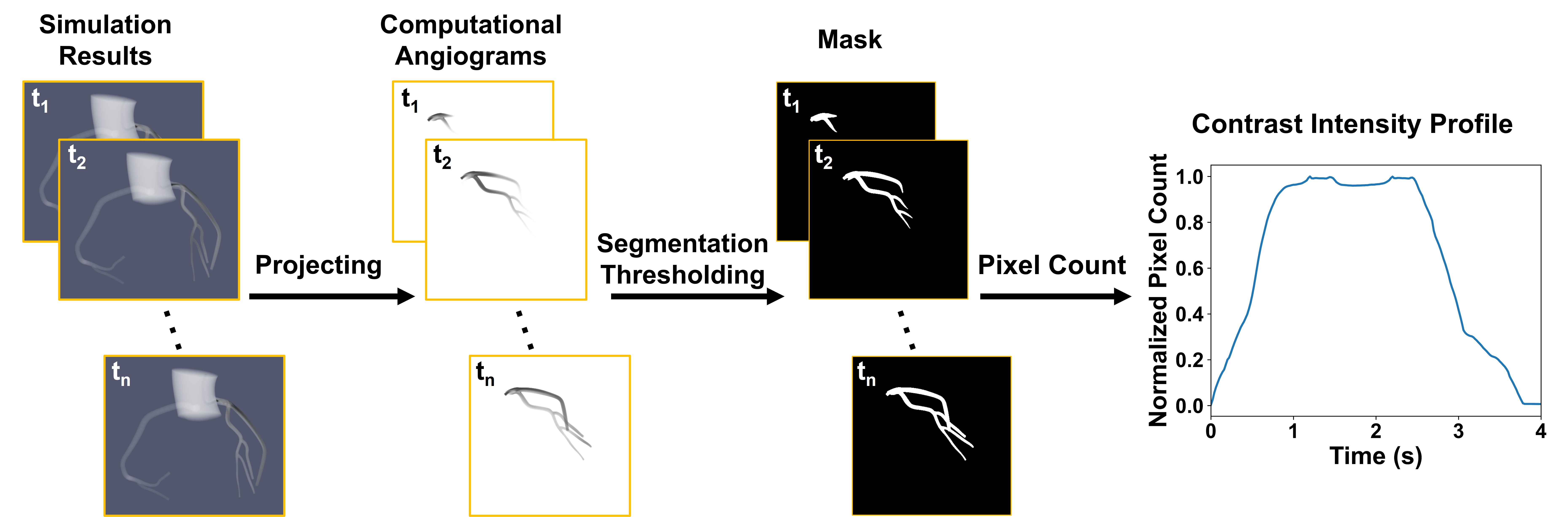}
\caption{\centering Extraction of computational CIP from multi-physics results.}
\label{fig:Computational_CIP_Generation}
\end{figure}

\underline{Post-processing - computational CIP extraction:} Once 3D solutions for velocity $\boldsymbol{u}$, pressure $p$, and iodine contrast concentration $c$ are obtained from the multi-physics simulations, computational CIPs can be extracted. Since clinical CIPs are extracted from 2D angiograms acquired under certain view angles, we must produce 2D computational angiograms from the 3D simulation results. Fig. \ref{fig:Computational_CIP_Generation} illustrates this process. First, 3D simulation results of iodine contrast concentration $c$ within the left coronary tree are projected onto a 2D plane using view angles similar to clinical X-ray angiograms: e.g., RAO (right anterior oblique), CRA (cranial), LAO (left anterior oblique), and CAU (caudal). A linear mapping between contrast concentration $c$ in the $(0, 400)$ $\text{mg/ml}$ scale and a $(0,255)$ grayscale is made. This projection and mapping is done for each frame of the simulation results, producing a series of 2D computational angiograms.
Then, segmentation of the computational angiograms is performed via thresholding with an arbitrary value of ($I_\text{thr}$ = 250), producing a binary mask for each angiogram. This operation assigns a value of 255 to pixels with intensity values less than 250 and 0 to those equal to or greater than 250, thereby producing a binary mask for each angiogram. The value of $I_\text{thr}$ was iteratively selected to produce good results for both resting and hyperemic conditions. Philosophically, this arbitrary threshold also exists when performing segmentation of clinical angiograms using a machine learning model \cite{iyer2021angionet,stevens2021,resnick2024neural}: the manual labels used to train the network also require the specification of an arbitrary threshold. 
Finally, the number of pixels within the binary segmentation mask is counted for each frame, and a normalized CIP is generated, representing the normalized pixel count over time.

%Similarly, contrast intensity profiles for clinical angiography data are extracted using a similar process, as illustrated in Fig. \ref{fig:Clinical_CIP_Generation}. First, patient-specific time-series angiograms are generated through cardiac catheterization, during which contrast is injected and X-ray images are taken to see how contrast moves through the coronary artery. These time-series angiograms are then processed using AngioNet, introduced in our previous study \cite{iyer2021angionet}, to segment the portions of the vessels filled with iodine contrast and create binary images. Finally, the number of bright pixels in the segmented images is counted for each frame, producing an intensity profile that reflects the injection and washout of the contrast agent.

% \begin{figure}[h]
% \centering
% \includegraphics[width=0.7\textwidth]{Clinical_CIP_Generation.jpg}
% \caption{\centering Contrast intensity profile generation from clinical angiography data.}
% \label{fig:Clinical_CIP_Generation}
% \end{figure}

\subsection{Model Calibration}\label{Model_Calibration}

To generate physiologically meaningful results, the multi-physics model of contrast injection requires calibration of its 43 adjustable parameters within the heart, aorta, and coronary LPMs, using the available patient data and additional literature data \cite{arthurs2016mathematical,casas2018non,rongen1999effect,fuchs2020high,bush1989effect,uribe2007whole,tarkin2013hemodynamic,biaggioni1987cardiovascular,stanojevic2016intravenous}. Hat quantities refer to patient-specific data. Non-hat quantities refer to simulation results. Calibration of these parameters takes place in two stages: 1) a \textit{hemodynamic calibration} of heart and aortic LPM parameters and 2) a \textit{coronary LPM calibration} using the angiography data. Details are provided next. 

\underline{Data:} We assumed knowledge of mean and peak cardiac outputs ($\hat{Q}_{\text{mean}}$ and $\hat{Q}_{\text{max}}$), systolic and diastolic blood pressure ($\hat{P}_{\text{sys}}$ and $\hat{P}_{\text{dia}}$), as well as lower bound and upper bound left ventricular end-diastolic volume ($\hat{\text{EDV}}_{\text{LB}}$ and $\hat{\text{EDV}}_{\text{UB}}$) \cite{biaggioni1987cardiovascular,bush1989effect,tarkin2013hemodynamic,stanojevic2016intravenous}. Additionally, patient-specific angiograms and their corresponding CIPs ($\hat{\text{CIP}}$) under both rest and hyperemia were obtained, along with invasive CFR measurements ($\hat{\text{CFR}}$). Table \ref{tab:TargetValues} summarizes the data for rest and hyperemia which will be used in the two-stage model calibration. 

\begin{table}[H]
\centering
\footnotesize
\renewcommand{\arraystretch}{1.5} % Adjusting row spacing
\setlength{\tabcolsep}{5pt}
\caption{\centering Patient-specific data for model calibration under both resting and hyperemic conditions}
\label{tab:TargetValues}
\begin{tabular}{c c c c c c c c c}
\hline
  & $\hat{Q}_{\text{mean}}$ & $\hat{Q}_{\text{max}}$ & $\hat{P}_{\text{sys}}$  & $\hat{P}_{\text{dia}}$ & $\hat{\text{EDV}}_{\text{UB}}$ & $\hat{\text{EDV}}_{\text{LB}}$ & Angiograms & $\hat{\text{CFR}}$ \\ 
  & (L/min) & (L/min) & (mmHg) & (mmHg) & (ml) & (ml) & &\\
\hline
Resting & 4.9 & 28 & 130 & 80 & 160 & 100 & $\hat{\text{CIP}}_{\text{rest}}$ & \multirow{2}{*}{2.2}\\ 
Hyperemia & 8.8 & 39 & 125 & 75 & 170 & 100 & $\hat{\text{CIP}}_{\text{hyper}}$ & \\ 
\hline
\end{tabular}
\end{table}

\underline{Stage 1 - Hemodynamic calibration:} Heart and aorta LPMs (including the elastance function) are calibrated to match patient-specific $\hat{Q}_{\text{mean}}$, $\hat{Q}_{\text{max}}$, $\hat{P}_{\text{sys}}$ and $\hat{P}_{\text{dia}}$ in the aorta. The objective function $L_1$ for this stage can be defined as:

\begin{equation}
\label{eqn:lf1}
\begin{aligned}
    & \underset{\textbf{x}}{\text{minimize}}
    & & L_{1} = \frac{\left|Q_{\text{mean}}-\hat{Q}_{\text{mean}}\right|}{\hat{Q}_{\text{mean}}}
+ \frac{\left|Q_{\text{max}}-\hat{Q}_{\text{max}}\right|}{\hat{Q}_{\text{max}}} 
+\frac{\left|P_{\text{sys}}-\hat{P}_{\text{sys}}\right|}{\hat{P}_{\text{sys}}}
+\frac{\left|P_{\text{dia}}-\hat{P}_{\text{dia}}\right|}{\hat{P}_{\text{dia}}} 
\\
    & \text{subject to}
    & & \hat{\text{EDV}}_{\text{LB}} <\text{EDV}<\hat{\text{EDV}}_{\text{UB}}\\
    & & & P_{\text{sys}}-0.5P_{\text{pulse}} <P_{\text{DN}}<P_{\text{sys}}-0.2P_{\text{pulse}}
\end{aligned}
\end{equation}
Here, $\textbf{x}$ is the vector of design variables, $\text{EDV}$ represents the end-diastolic volume, 
$P_\text{DN}$ the dicrotic notch pressure, which is the pressure in the arteries when the aortic valve closes, and $P_{\text{pulse}} = P_{\text{sys}}-P_{\text{dia}}$ is the pulse pressure. $\text{EDV}$ is constrained to remain within its lower and upper bounds ($\hat{\text{EDV}}_{\text{LB}}$ and $\hat{\text{EDV}}_{\text{UB}}$) during optimization. Furthermore, the dicrotic notch pressure $P_\text{DN}$ is constrained to remain within a certain range of the pulse pressure, to achieve physiologically realistic pressure waveforms. 

The design variables vector $\textbf{x}$ comprises seven parameters within the heart and aorta LPMs: systemic compliance $C_\text{s}$, proximal systemic resistance $R_\text{sp}$, distal systemic resistance $R_\text{sd}$, maximum elastance $E_\text{max}$, minimum elastance $E_\text{min}$, time to maximum elastance $t_\text{max}$, and left atrial pressure $P_\text{LA}$, see Table \ref{tab:rangeDV}. Ranges for each design variable under both resting and hyperemia conditions, based on literature \cite{arthurs2016mathematical,garg2023acute,casas2018non,rongen1999effect,fuchs2020high,tarkin2013hemodynamic,dobson2017adenosine,biaggioni1987cardiovascular,stanojevic2016intravenous}, is also provided in the table. The asterisk (*) represents the optimal design parameters, while the superscript `r' indicates the resting condition. The range of the design variables during hyperemia is derived from their value under the resting condition.

\begin{table}[H]
\centering
%\scriptsize
\renewcommand{\arraystretch}{1.5} % Adjusting row spacing
\setlength{\tabcolsep}{0.5pt}
\caption{\centering Parameters of aortic and heart LPM which are optimized during calibration, making up the design variables vector $\textbf{x}$. Brackets provide the specified range of design variables} 
\label{tab:rangeDV}
\resizebox{\textwidth}{!}{  % Automatically resize the table to fit within the text width
\begin{tabular}{c c c c c c c c}
\hline
  & $C_\text{s}$ & $R_\text{sp}$ & $R_\text{sd}$ & $E_\text{max}$ & $E_\text{min}$ & $t_\text{max}$ & $P_\text{LA}$ \\ 
  & ($\text{mm}^3 \cdot \text{Pa}^{-1}$) & ($\text{Pa} \cdot \text{s} \cdot \text{mm}^{-3}$) & ($\text{Pa} \cdot \text{s} \cdot \text{mm}^{-3}$) & ($\text{Pa} \cdot \text{mm}^{-3}$) & ($\text{Pa} \cdot \text{mm}^{-3}$) & (s) & (Pa) \\
\hline
$\text{Range}_{\text{Resting}}$ & [5, 30] & [0.003, 0.015] & [0.075, 0.3] & [0.133, 0.266] & [0.0053, 0.016] & [0.35, 0.39] & [1500, 2300]   \\ 
$\text{Range}_{\text{Hyper}}$ & [$C_\text{s}^\text{r*}$-15, $C_\text{s}^\text{r*}$-2] & [0.3, 0.9]$R_\text{sp}^\text{r*}$ & [0.3, 0.9]$R_\text{sd}^\text{r*}$ & [1.2$E_\text{max}^\text{r*}$, 0.533] & $E_\text{min}^\text{r*}$ & [0.95,1.05]$t_\text{max}^\text{r*}$ & [1.02, 1.08] $P_\text{LA}^\text{r*}$  \\ 
\hline
\end{tabular}
}
\end{table}

The remaining heart LPM parameters were held constant  \cite{arthurs2016mathematical} under both hemodynamics conditions, as detailed in Table \ref{tab:Heart_parameters}. This simplification is justified because these parameters have minimal influence on the objective function, or can be derived from patient-specific data. For example, the cardiac cycle $T$ was obtained from patient-specific angiograms.
\begin{table}[H]
\centering
\footnotesize
\renewcommand{\arraystretch}{1.5} % Adjusting row spacing
\setlength{\tabcolsep}{5pt}
\caption{\centering Parameters of heart LPM and elastance function which are kept fixed during calibration}
\label{tab:Heart_parameters}
\begin{tabular}{c c c c c c c}

\hline
  $R_{\text{MV}}$ & $L_{\text{MV}}$ & $R_{\text{AV}}$ & $L_{\text{AV}}$ & $t_\text{r} - t_\text{max}$ & $T_{\text{Rest}}$ & $T_{\text{Hyper}}$ \\ 
  ($\text{Pa} \cdot \text{s} \cdot \text{mm}^{-3}$) & ($\text{Pa} \cdot \text{s}^2 \cdot \text{mm}^{-3}$) & ($\text{Pa} \cdot \text{s} \cdot \text{mm}^{-3}$) & ($\text{Pa} \cdot \text{s}^2 \cdot \text{mm}^{-3}$) & (s) & (s) & (s) \\
\hline
  $3.9 \times 10^{-4}$ & $1 \times 10^{-5}$ & $1 \times 10^{-5}$ & $1 \times 10^{-5}$ & 0.1 & 1 & 0.73 \\
\hline
\end{tabular}
\end{table}
To ensure computational efficiency in the calibration process, the aorta and coronary arteries are replaced by resistors, effectively transforming the 3D-0D model into a 0D model which can be run nearly instantaneously \cite{arthurs2021crimson}. Despite this simplification, these `pure 0D' simulations provide reasonably accurate flow and pressure waveforms at a fraction of the computational cost of the 3D-0D model. The optimization is carried out using a differential evolution algorithm \cite{storn1997differential} with 49 individuals per population and a maximum of 150 generations. Optimization continues until either the maximum number of generations is reached or the population’s standard deviation falls below the tolerance level, defined as $0.01\times L^\text{mean}_1$, where $L^\text{mean}_1$ is the average loss of the current generation. Once the heart and aorta LPM parameters are optimized to achieve the target hemodynamics, the process moves on to the second stage of parameter tuning for the coronary model.

\underline{Stage 2 - coronary LPM calibration}: In the second stage, coronary artery models are tuned to match patient-specific CIPs ($\hat{\text{CIP}}$) under resting and hyperemic conditions, as well as the invasive CFR measurement $\hat{\text{CFR}}$. The objective function for this stage can be expressed as:
\begin{equation}
\label{eqn:lf2}
\begin{aligned}
    & \underset{\textbf{x}}{\text{minimize}}
    & & L_2 = \|\text{CIP}_{\text{rest}}-\hat{\text{CIP}}_{\text{rest}}\|_2%\frac{ \|\text{CIP}_{\text{rest}}-\hat{\text{CIP}}_{\text{rest}}\|_2}{\|\hat{\text{CIP}}_{\text{mean}}\|_2}
+ \|\text{CIP}_{\text{hyper}}-\hat{\text{CIP}}_{\text{hyper}}\|_2%\frac{\left|\text{CIP}_{\text{hyper}}-\hat{\text{CIP}}_{\text{hyper}}\right|}{\hat{\text{CIP}}_{\text{hyper}}}
+\frac{\left|\text{CFR}-\hat{\text{CFR}}\right|}{\hat{\text{CFR}}} 
\end{aligned}
\end{equation}
Here, $\text{CIP}_\text{rest}$ and $\text{CIP}_\text{hyper}$ represent the computational CIPs under resting and hyperemic conditions, respectively, while $\text{CFR}$ is the value derived from simulations of both states. Even though we only consider contrast intensity profiles derived from angiograms in the left coronary tree ($\hat{\text{CIP}}$) and their computational counterparts (CIP), in this paper we estimate coronary LPM parameters for both left and right coronary trees, for a total of six branches (see Fig. \ref{fig:CFD_model}). Fundamentally, this assumption implies that similar degrees of CMD are present on both sides. Therefore, the design variables vector \textbf{x} consists of 30 parameters (6 vessels, 5 parameters per vessel) in the left and right coronary LPMs: large artery resistance $R_{\text{a},i}$, proximal small artery resistance $R_{\text{ap},i}$, distal small artery resistance $R_{\text{ad},i}$, arterial compliance $C_{\text{a},i}$, and intramyocardial compliance $C_{\text{im},i}$, for $i = 1,2,...,6$. 

Evaluating the objective function $L_2$ (Eq. \ref{eqn:lf2}) for each parameter set $\textbf{x}$ requires two computationally intensive 3D-0D multi-physics simulations, each taking approximately 100 hours to complete on a 108-core computing system. Therefore, direct optimization is not desired. Instead, a two-step tuning strategy was implemented, as outlined in Algorithm \ref{alg:coronary-calibration}: (1) A 0D pre-tuning step, whereby coronary artery LPMs were pre-tuned by iteratively running 0D simulations to achieve flow splits based on Murray’s Law \cite{taylor2024systematic}: $r_0^3 = r_1^3 + r_2^3$ for proximal branch of radius $r_0$ splitting into distal branches with radii $r_1,r_2$, and match target waveforms in the coronary arteries \cite{hozumi1998noninvasive} under both resting and hyperemic conditions. (2) A grid search step to identify the optimal combination of parameters for both rest and hyperemia that minimized Eq. \ref{eqn:lf2}.

During the 0D pre-tuning step, the design variables $\textbf{x}$ were iteratively adjusted to match flow splits derived from Murray’s Law, where the flow rate $Q_i$ at branch $i$ is proportional to $r_i^3$. Additionally, flow waveforms in each coronary artery were tuned to exhibit a diastolic-dominant profile, with greater dominance in the left coronary tree compared to the right. For example, both $R_\text{ap}$ and $C_\text{a}$ influence the diastolic flow dominance, and $R_\text{ad}$ regulates the magnitude of backflow, etc. In the Grid search step, seven 3D-0D multi-physics simulations were performed by perturbing the pre-tuned total resistance for each coronary artery about each state: $R_{\text{T},i} = R_{\text{a},i} + R_{\text{ap},i} + R_{\text{ad},i}$ by $\pm3$\% per simulation, resulting in 49 total parameter combinations. The total compliance for each coronary artery 
$C_{\text{T},i} = C_{\text{a},i} + C_{\text{im},i}$
was kept constant during the grid search. 

The combination of parameters that minimizes the objective function $L_2$ (Eq. \ref{eqn:lf2}) defines the final fine-tuned multi-physics models for rest and hyperemia. This approach produces computational CIPs that closely align with clinical CIPs while accurately reproducing invasive CFR measurements. Throughout the coronary LPM calibration stage, heart and aorta LPM parameters obtained in the hemodynamic calibration stage were held constant, as changes in the coronary artery LPMs, accounting for only a small percentage of the cardiac output, have minimal impact on systemic hemodynamics. 

\begin{algorithm}
   \caption{Coronary LPM Calibration Algorithm}
   \label{alg:coronary-calibration}
    \begin{algorithmic}[1]
     \State\textbf{\textcolor{red}{0D Pre-tuning:}}
        \For{$i = 1$ to $2$} \Comment{$1$ and $2$ represent resting and hyperemic conditions, respectively}
            \State Initialize $\textbf{x}$ with an initial guess \cite{arthurs2016mathematical}
            \While{convergence criterion not met}
                \State Perform simulations iteratively, updating \textbf{x} to match:
                \begin{itemize}
                    \item Flow splits based on Murray's Law
                    \item Target diastolic-dominant flow waveforms in coronary arteries
               \end{itemize}
            \EndWhile
        \EndFor
        
    \State\textbf{\textcolor{red}{Grid Search:}}
        \State Perform multi-physics simulations with small perturbations ($\pm$3\%) around the pre-tuned branch resistances $R_{\text{T},i}$ for each state
        \State Identify the parameter combination across both states that minimizes $L_{2}$ 
    \end{algorithmic}
\end{algorithm}

\section{Results}\label{Results}

In this section, we discuss two different sets of results. In the first set, calibration of the model under both resting and hyperemic conditions is demonstrated by matching targeted hemodynamics and angiographic data. In the second set, we conduct different sensitivity studies to understand the correlation between the parameters of the coronary LPM and features in the CIP. 

\subsection{Calibration of Multi-physics Model}\label{result_cali}

Following the process outlined in section \ref{Model_Calibration}, calibration of the LPM parameters of the model under both resting and hyperemic conditions is demonstrated by matching targeted hemodynamics and angiographic data. 

\begin{figure}[H]
\centering
\includegraphics[width=1\textwidth]{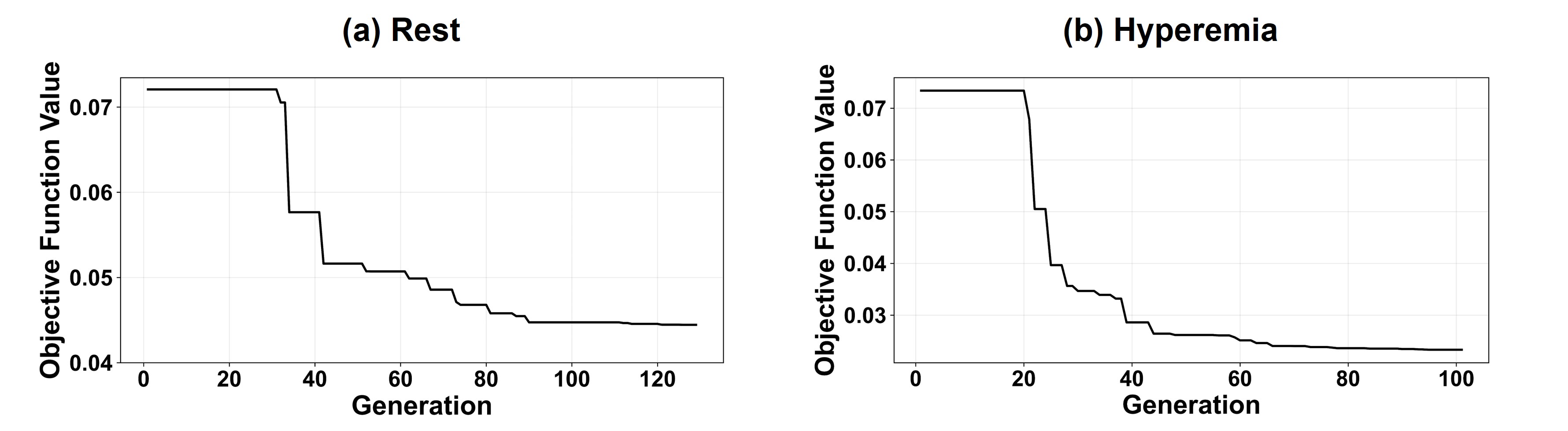}
\caption{\centering Convergence curves of differential evolution optimization for the objective function $L_1$ (Eq. \ref{eqn:lf1}) under (a) rest and (b) hyperemia.}
\label{fig:convergence}
\end{figure}

\underline{Stage 1 Calibration:} Heart and aorta models were calibrated to minimize Eq. \ref{eqn:lf1}. Fig. \ref{fig:convergence} shows convergence curves of differential evolution optimization for the objective function $L_1$ under (a) rest and (b) hyperemia. The optimal designs for both conditions are presented in Table \ref{tab:Optimized_Heart_parameters}. The convergence curves show that the optimization for both cases stops because the standard deviation of the current population falls below the specified threshold. Results show that, under hyperemia, systemic compliance $C_\text{s}$, proximal systemic resistance $R_\text{sp}$, and distal systemic resistance $R_\text{sd}$ all decreased by 11.0\%, 44.4\%, and 44.9\%, respectively, compared to resting conditions. Maximum elastance $E_\text{max}$ increased by 20.0\%. Additionally, both time to maximum elastance $t_\text{max}$ and left atrial pressure $P_\text{LA}$ increased by 4.9\% and 3.9\%, respectively.

% \begin{table}[H]
% \centering
% \footnotesize
% \renewcommand{\arraystretch}{1.5} % Adjusting row spacing
% \setlength{\tabcolsep}{5pt}
% \caption{\centering Target values for objective variables and constraints for both resting and hyperemic conditions}
% \label{tab:TargetValues}
% \begin{tabular}{c c c c c c c}
% \hline
%   & $\hat{Q}_{\text{mean}}$ & $\hat{Q}_{\text{max}}$ & $\hat{P}_{\text{sys}}$  & $\hat{P}_{\text{dia}}$ & $\text{EDV}_{\text{LB}}$ & $\text{EDV}_{\text{UB}}$ \\ 
%   & (L/min) & (L/min) & (mmHg) & (mmHg) & (ml) & (ml) \\
% \hline
% Resting & 4.9 & 28 & 130 & 80 & 160 & 100 \\ 
% Hyperemia & 8.8 & 39 & 125 & 75 & 170 & 100 \\ 
% \hline
% \end{tabular}
% \end{table}

\begin{table}[H]
\centering
%\scriptsize
\footnotesize

\renewcommand{\arraystretch}{1.5} % Adjusting row spacing
\setlength{\tabcolsep}{5pt}
\caption{\centering The optimal designs for both resting and hyperemic conditions}
\label{tab:Optimized_Heart_parameters}
%\resizebox{\textwidth}{!}{  % Automatically resize the table to fit within the text width
\begin{tabular}{c c c c c c c c}
\hline
  & $C_\text{s}$ & $R_\text{sp}$ & $R_\text{sd}$ & $E_\text{max}$ & $E_\text{min}$ & $t_\text{max}$ & $P_\text{LA}$ \\ 
  & ($\text{mm}^3 \cdot \text{Pa}^{-1}$) & ($\text{Pa} \cdot \text{s} \cdot \text{mm}^{-3}$) & ($\text{Pa} \cdot \text{s} \cdot \text{mm}^{-3}$) & ($\text{Pa} \cdot \text{mm}^{-3}$) & ($\text{Pa} \cdot \text{mm}^{-3}$) & (s) & (Pa) \\
\hline

$\textbf{x}^\text{r*}$ & 18.381 & 0.009 & 0.158 & 0.190 & 0.015 & 0.390 & 2286.880  \\ 

$\textbf{x}^{h*}$ & 16.361 & 0.005 & 0.087 & 0.228 & 0.015 & 0.409 & 2376.910  \\ 

\hline
\end{tabular}
%}
\end{table}

% \begin{table}[H]
% \centering
% \footnotesize
% \renewcommand{\arraystretch}{1.5} % Adjusting row spacing
% \setlength{\tabcolsep}{5pt}
% \caption{\centering Other parameters in the heart model and the elastance function}
% \label{tab:Heart_parameters}
% \begin{tabular}{c c c c c c c}

% \hline
%   $R_{\text{MV}}$ & $L_{\text{MV}}$ & $R_{\text{AV}}$ & $L_{\text{AV}}$ & $t_\text{r} - t_\text{max}$ & $T_{\text{Rest}}$ & $T_{\text{Hyper}}$ \\ 
%   ($\text{Pa} \cdot \text{s} \cdot \text{mm}^{-3}$) & ($\text{Pa} \cdot \text{s}^2 \cdot \text{mm}^{-3}$) & ($\text{Pa} \cdot \text{s} \cdot \text{mm}^{-3}$) & ($\text{Pa} \cdot \text{s}^2 \cdot \text{mm}^{-3}$) & (s) & (s) & (s) \\
% \hline
%   $3.9 \times 10^{-4}$ & $1 \times 10^{-5}$ & $1 \times 10^{-5}$ & $1 \times 10^{-5}$ & 0.1 & 1 & 0.73 \\
% \hline
% \end{tabular}
% \end{table}

\begin{figure}[H]
\centering
\includegraphics[width=1\textwidth]{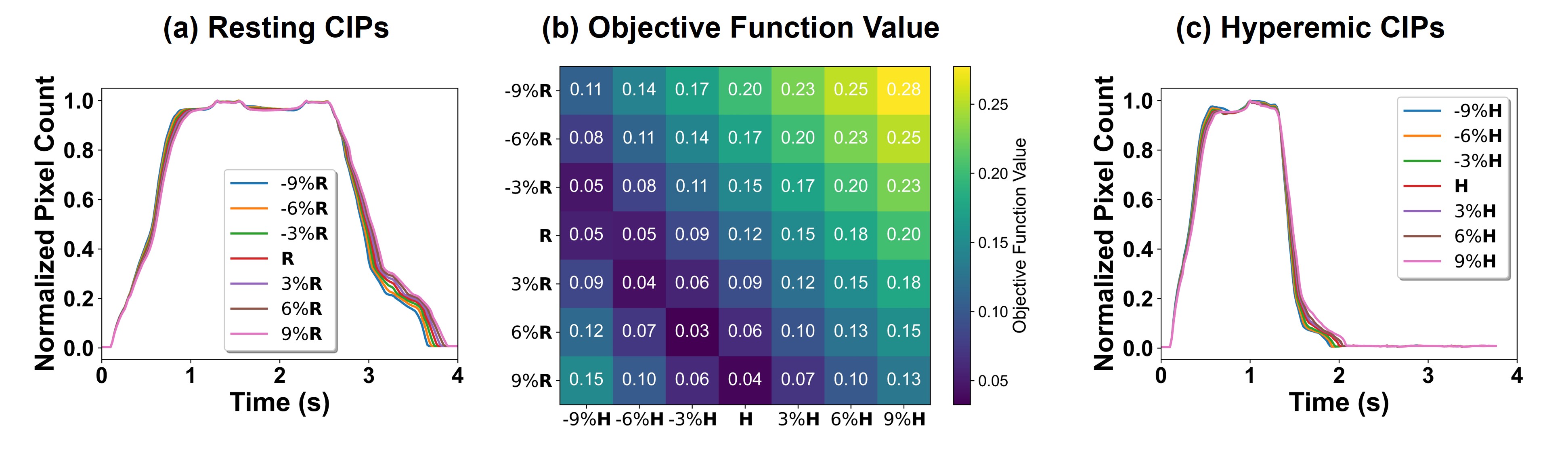}
\caption{\centering CIPs under (a) resting and (c) hyperemic conditions during the grid search, along with (b) the objective function values corresponding to the grid search.}
\label{fig:ObjFuncV_table}
\end{figure}

\underline{Stage 2 Calibration:} 
Calibration of the coronary LPMs is performed according to Algorithm \ref{alg:coronary-calibration}. The 0D Pre-tuning stage produced reference parameter sets, which we denote as \textbf{R} and \textbf{H}. The Grid Search stage is performed by disturbing the total resistances for each reference parameter set \textbf{R} ($R_{\text{T},i}^\text{r}$) and \textbf{H} ($R_{\text{T},i}^\text{h}$) by $\pm 3 \%, \pm 6 \%$, and $\pm 9 \%$, to define a grid combining solutions for resting and hyperemic stages. Fig. \ref{fig:ObjFuncV_table} (b) illustrates values of the objective function $L_{2}$ for all instances of the Grid Search, combining variations in resting and hyperemic conditions.
Fig. \ref{fig:ObjFuncV_table} (a), and (c) show the CIPs for all instances of the grid search rest and hyperemia, respectively. 
The combination of resting and hyperemic solutions which minizime $L_{2}$ is (6\%R \& -3\%H), which denotes an increase of $6\%$ of the resistances ($R_{\text{T},i}^\text{r}$) from the reference rest parameter set \textbf{R} and a decrease of $3\%$ of resistances 
($R_{\text{T},i}^\text{h}$) of the reference hyperemia parameter set \textbf{H}. This optimal set of parameters is given in Table \ref{tab:coronary_parameters}. Results show that the average decrease from rest to hyperemia of resistances and capacitances of the coronary LPM is 56.6\% and 5.0\%, respectively.

\begin{figure}[H]
\centering
\includegraphics[width=1\textwidth]{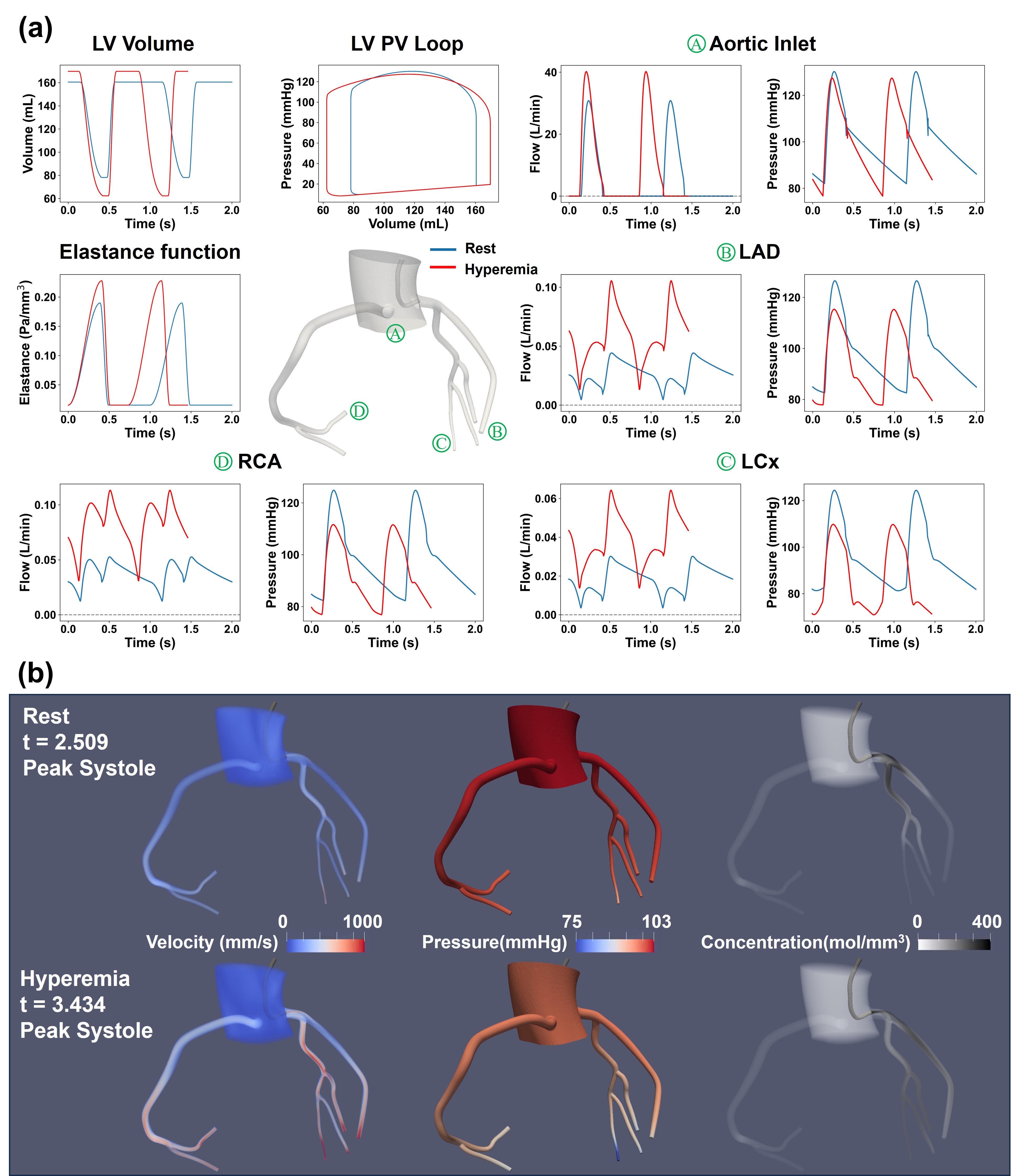}
\caption{\centering 3D-0D coupled multi-physics CFD simulation results: (a) cardiac function and coronary artery hemodynamics for resting and hyperemic conditions. (b) CFD contour plots of velocity, pressure, and contrast concentration at peak systole for rest (top row) and hyperemia (bottom row).}
\label{fig:Hemodynamics}
\end{figure}

\underline{Calibrated hemodynamics:} Fig. \ref{fig:Hemodynamics} displays the results of the 3D-0D coupled multi-physics CFD simulations calibrated for both resting and hyperemic conditions. Panel (a) illustrates the cardiac function hemodynamics, including left ventricular (LV) volume, elastance function, and pressure-volume (PV) loop, along with the flow and pressure waveforms at the aortic inlet, LAD, LCx, and RCA. Panel (b) shows velocity, pressure, and contrast concentration contour plots at peak systolic flows, $t_\text{rest} = 2.509$ s for rest, and $t_\text{hyper} = 3.434$ s for hyperemia. Coronary artery flow waveforms display physiologically realistic patterns, namely a diastolic-dominant flow for both resting and hyperemic conditions. The intramyocardial pressure $P_\text{im}$, integrated into the coronary model (see Fig. \ref{fig:CFD_model}), contributes to the distinct diastolic-dominant coronary flow patterns depicted in Fig. \ref{fig:Hemodynamics}. The left ventricle contracts with greater force, generating larger ventricular and intramyocardial pressures on the left coronary circulation compared to the right.

\begin{table}[H]
\centering
\footnotesize
\renewcommand{\arraystretch}{1.5} % Adjusting row spacing
\setlength{\tabcolsep}{5pt}
\caption{\centering Parameter values in coronary artery model}
\label{tab:coronary_parameters}
\begin{tabular}{c c c c c c c}
\hline
  &  & $R_\text{a}$ & $R_\text{ap}$ & $R_\text{ad}$ & $C_\text{a}$ & $C_\text{im}$ \\ 
  &  & ($\text{Pa} \cdot \text{s} \cdot \text{mm}^{-3}$) & ($\text{Pa} \cdot \text{s} \cdot \text{mm}^{-3}$) & ($\text{Pa} \cdot \text{s} \cdot \text{mm}^{-3}$) & ($\text{mm}^3 \cdot \text{Pa}^{-1}$) & ($\text{mm}^3 \cdot \text{Pa}^{-1}$) \\
\hline
     \multirow{6}{*}{Rest}& LAD &7.130& 2.139& 19.923& 0.014& 0.135
\\ 
     & OM1 &5.857& 1.757& 16.366& 0.014& 0.135
\\ 
     & OM2 &11.225& 3.368& 31.367& 0.007& 0.135
\\
     & LCx &10.040& 3.012& 28.055& 0.007& 0.135
\\
     & AM &7.465& 2.239& 20.860& 0.012& 0.189
\\
     & RCA &5.020& 1.506& 14.028& 0.012& 0.189
\\ 
\hline
     \multirow{6}{*}{Hyperemia}& LAD &3.091& 0.927& 8.637& 0.014& 0.128
\\ 
     & OM1 &2.539& 0.762& 7.095& 0.014& 0.128
\\
     & OM2 &4.866& 1.460& 13.598& 0.007& 0.128
\\
     & LCx &4.352& 1.306& 12.162& 0.007& 0.128
\\
     & AM &3.236& 0.971& 9.043& 0.011& 0.180
\\
     & RCA &2.176& 0.653& 6.081& 0.011& 0.180
\\
     
\hline
\end{tabular}

\end{table}

Results for key hemodynamic indices are summarized in Table \ref{tab:Hemo_numerical} (see the comparison of some of those indices with their target values given in Table \ref{tab:TargetValues}). Results indicate a 9.277 ml increase in end-diastolic volume ($\text{EDV}$), a 15.824 ml decrease in end-systolic volume ($\text{ESV}$), a 25.101 ml increase in stroke volume, and a 12.0\% rise in ejection fraction from resting to hyperemic conditions. This led to a 78.7\% increase in cardiac output ($Q_{\text{mean}}$) and a reduction of 2.770 and 5.363 mmHg in systolic ($P_{\text{sys}}$) and diastolic ($P_{\text{dia}}$) aortic pressures, respectively. Furthermore, the ratio of flow rates in the LAD under both resting and hyperemic conditions produce a CFR of 2.203, closely aligning with the clinical measurement of $\hat{\text{CFR}} = 2.2$, demonstrating that the model calibration is physiologically meaningful for both resting and hyperemic conditions.

\begin{table}[H]
\centering
\footnotesize
\renewcommand{\arraystretch}{1.5} % Adjusting row spacing
\setlength{\tabcolsep}{5pt}
\caption{\centering Hemodynamic metrics from calibrated simulations under resting and hyperemic conditions}
\label{tab:Hemo_numerical}
\resizebox{\textwidth}{!}{  % Automatically resize the table to fit within the text width
\begin{tabular}{c c c c c c c c c c c}
\hline
  & $Q_{\text{mean}}$ & $Q_{\text{max}}$ & $P_{\text{sys}}$  & $P_{\text{dia}}$ & $\text{EDV}$& $\text{ESV}$& SV & EF& $Q_\text{LAD}$ & CFR\\ 
  & (L/min) & (L/min) & (mmHg) & (mmHg) & (ml) & (ml)  & (ml) &  &(L/min)& \\
\hline
Rest & 4.941& 30.822& 130.249& 82.013& 160.517& 78.147& 82.370
& 51.315\%& 0.027 &  \multirow{2}{*}{2.203}\\ 
 Hyperemia & 8.830
& 40.206& 127.480& 76.650& 169.794& 62.323& 107.471& 63.295\%& 0.059& \\ 
\hline
\end{tabular}
}
\end{table}

\begin{figure}[h]
\centering
\includegraphics[width=0.7\textwidth]{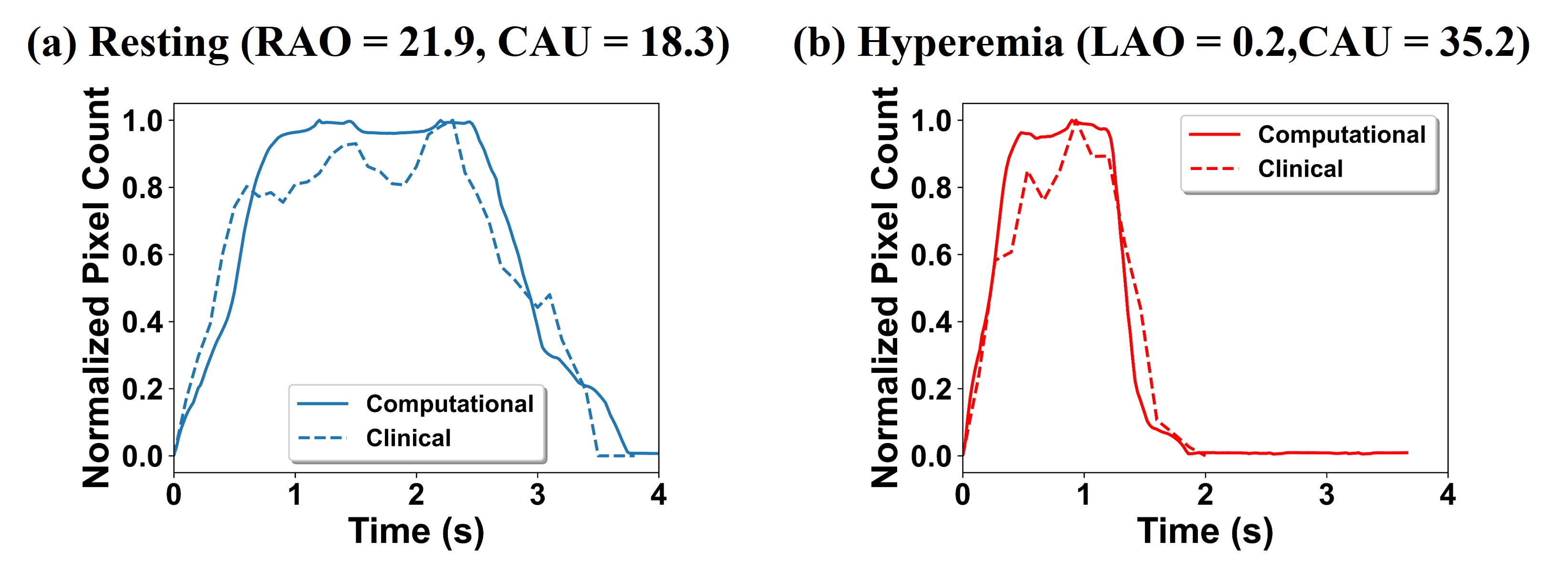}
\caption{\centering CIP comparison between computational and clinical data for resting and hyperemic conditions}
\label{fig:CIP_comparison}
\end{figure}

\begin{figure}[h]
\centering
\includegraphics[width=0.8\textwidth]{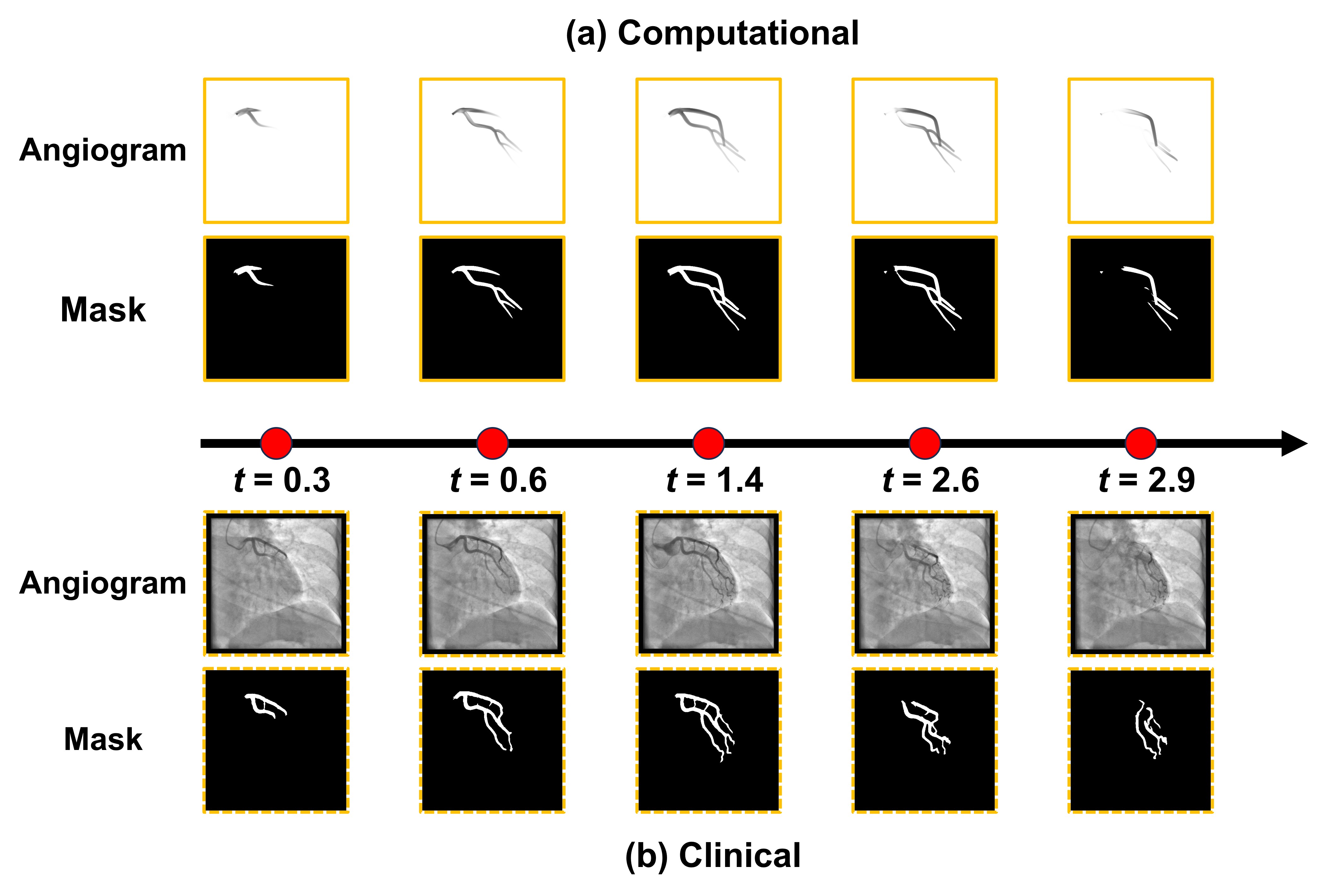}
\caption{\centering Angiogram comparison between computational and clinical data for resting condition}
\label{fig:Angios_comparison_resting}
\end{figure}

\begin{figure}[h]
\centering
\includegraphics[width=0.8\textwidth]{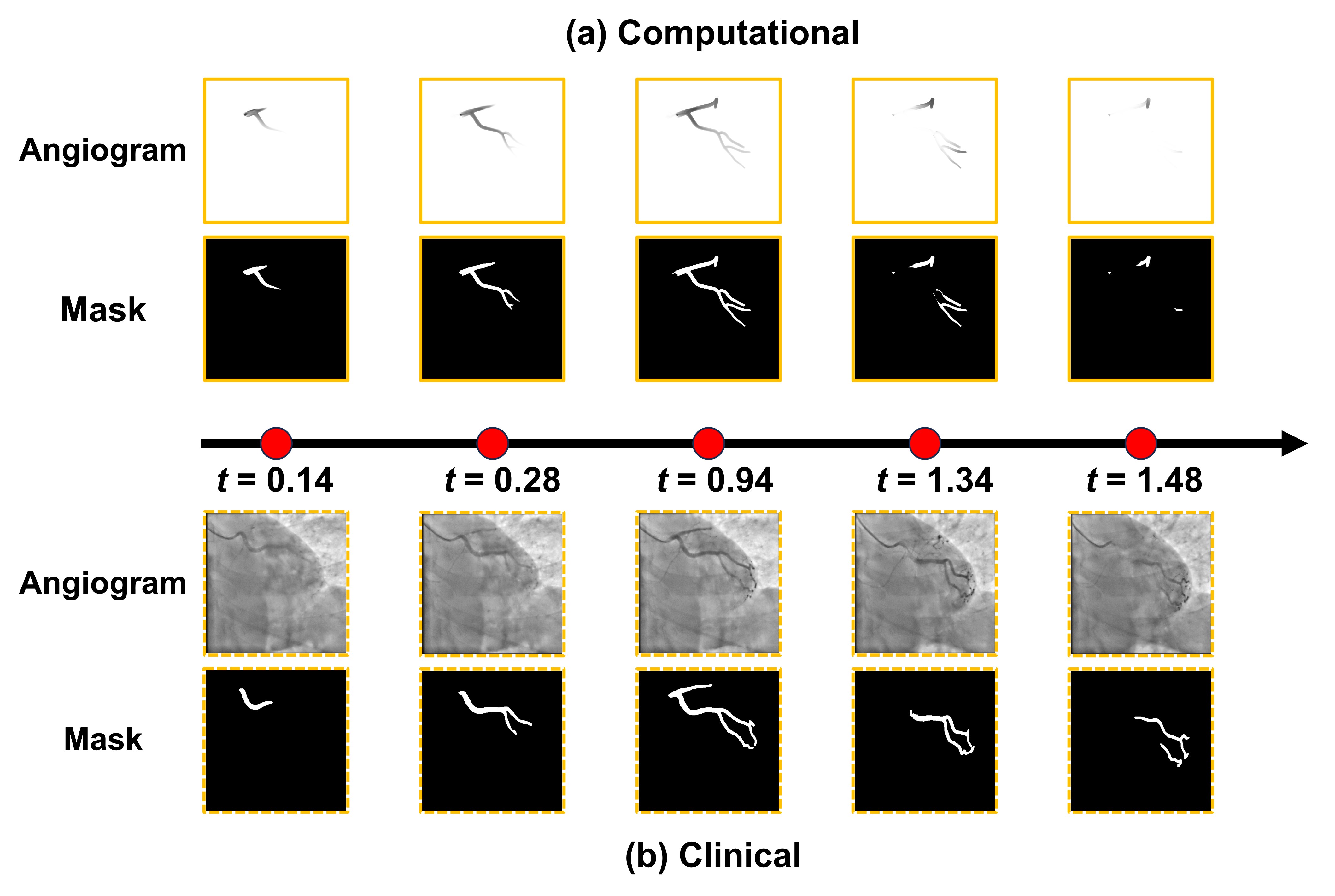}
\caption{\centering Angiogram comparison between computational and clinical data for hyperemic condition}
\label{fig:Angios_comparison_hyper}
\end{figure}

\underline{Calibrated CIPs:} Fig. \ref{fig:CIP_comparison} presents the comparison between computational and clinical CIPs under (a) resting conditions, under view angle (RAO = $21.9^\circ$, CAU = $18.3^\circ$) and (b) hyperemic conditions, under view angle (LAO = $0.2^\circ$, CAU = $35.2^\circ$). The results demonstrate that the calibrated computational CIPs generally show good agreement with the clinical CIPs, with mean squared errors of 0.0155 for rest and 0.0159 for hyperemia. Figs. \ref{fig:Angios_comparison_resting} and \ref{fig:Angios_comparison_hyper} show frame-by-frame angiogram comparisons at selected time points for both resting and hyperemic conditions. Overall, the computational angiograms and corresponding segmentation masks closely align with their clinical data counterparts throughout the injection and washout phases.

\subsection{Sensitivity studies} 

In this section, we conducted two different sensitivity studies. In the first study, perturbations were applied to individual parameters of the coronary LPMs, using a hyperemic state as the baseline. The goal of this study is to understand the impact that individual parameters have on features of the CIP. In the second study, perturbations were proportionally distributed across the different resistances of the coronary LPMs, using a resting state as the baseline. The goal of this study is to understand the impact of varying degrees of microvascular disease on features of the CIP.

\underline{Study 1: Perturbations on individual LPM parameters:} 
A healthy hyperemic condition was calibrated such that: cardiac output  $Q_{\text{mean}} = 9.008$ L/min, maximum cardiac output  $Q_{\text{max}} = 40.563$ L/min, systolic pressure $P_{\text{sys}} = 124.757$ mmHg, and diastolic pressure  $P_{\text{dia}} = 73.334$ mmHg.  The total left coronary flow in this state was 0.47 L/min (5.2 \% of the cardiac output). This defines a baseline condition (which we refer to as 1x) for all parameters.

Starting with the 1x baseline healthy hyperemic condition, each coronary LPM parameter for each branch $i$ was multiplied by a sequence of factors. The large artery resistance $R_{\text{a},i}$, proximal small artery resistance $R_{\text{ap},i}$, and distal small artery resistance $R_{\text{ad},i}$, were multiplied by factors $\{1\text{x}, 3\text{x}, 5\text{x}, 7\text{x}, 9\text{x}\}$. The arterial compliance $C_{\text{a},i}$ and intramyocardial compliance $C_{\text{im},i}$ were multiplied by factors $\{1\text{x}, 1/3\text{x}, 1/5\text{x}, 1/7\text{x}, 1/9\text{x}\}$. Each of these perturbations results in increased vascular resistance and decreased vascular compliance, both associated with increased degree of CMD.  

\begin{figure}[H]
\centering
\includegraphics[width=1\textwidth]{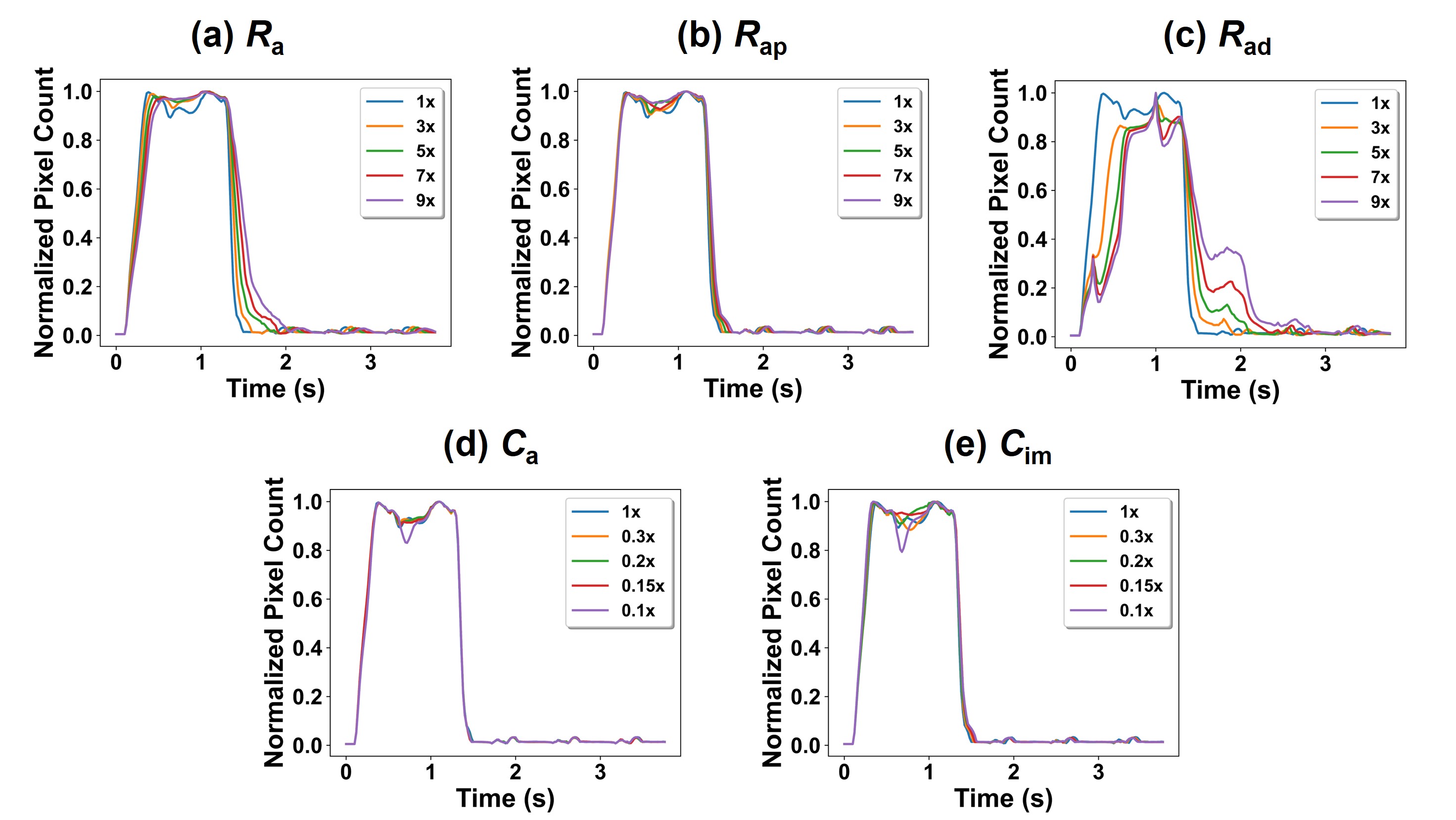}
\caption{\centering The impact of the coronary model parameters on CIP: (a) large artery resistance $R_\text{a}$, (b) proximal small artery resistance $R_\text{ap}$, (c) distal small artery resistance $R_\text{ad}$, (d) arterial compliance $C_\text{a}$, and (e) intramyocardial compliance $C_\text{im}$}
\label{fig:sensitivity_study}
\end{figure}

Fig. \ref{fig:sensitivity_study} illustrates the CIPs calculated for each of the parameter perturbations defined above. Simulations were run by only perturbing one parameter at a time, while keeping the rest at the 1x reference value. Table \ref{tab:Hemo_numerical_sensitivity_study} summarizes hemodynamic indices for all perturbations. Results indicate that resistance has a greater effect on CIP shape than capacitance, significantly affecting rising and falling slopes. Larger resistances result in shallower slopes, with more pronounced effects on the falling slopes. Notably, the distal small artery resistances ($R_{\text{ad},i}$) have the largest impact on the CIP slopes. In contrast, compliances have minimal impact on the slope of the CIP. The main impact of compliance was on the shape of pulsatile coronary flow waveforms (not shown here), which resulted in alterations in the plateau of the CIP (around t $\approx$ 0.7 s). Similar trends were observed for hemodynamic indices in Table \ref{tab:Hemo_numerical_sensitivity_study}. The distal small artery resistance $R_{\text{ad},i}$ had the largest impact on all quantities, while the compliances had almost no impact.

\begin{table}[H]
\centering
\footnotesize
\renewcommand{\arraystretch}{1.5} % Adjusting row spacing
\setlength{\tabcolsep}{5pt}
\caption{\centering Hemodynamic metrics of sensitivity study}
\label{tab:Hemo_numerical_sensitivity_study}
\resizebox{\textwidth}{!}{  % Automatically resize the table to fit within the text width
\begin{tabular}{c c c c c c c c c c c}
\hline
&  & $Q_{\text{mean}}$ & $Q_{\text{max}}$ & $P_{\text{sys}}$  & $P_{\text{dia}}$ & $\text{EDV}$& $\text{ESV}$& SV & EF& $Q_\text{left tree}$\\ 
&  & (L/min) & (L/min) & (mmHg) & (mmHg) & (ml) & (ml)  & (ml) &  &(L/min)\\
\hline
\multirow{5}{*}{(a)  $R_\text{a}$}&1x & 9.008& 40.572& 124.753& 73.335
& 170.077& 60.443& 109.633& 64.461\%& 0.471
\\ 
   &3x & 8.931& 40.413& 125.916& 74.745
& 169.953& 61.267& 108.687& 63.951\%& 0.355
\\ 
   &5x & 8.883& 40.313& 126.653& 75.639
& 169.878& 61.768& 108.110& 63.640\%& 0.283
\\ 
   &7x & 8.851& 40.241& 127.140& 76.230
& 169.827& 62.101& 107.726& 63.433\%& 0.235
\\ 
   &9x & 8.829& 40.186& 127.502& 76.671
& 169.791& 62.341& 107.451& 63.284\%& 0.200
\\ 
\hline
\multirow{5}{*}{(b) $R_\text{ap}$}&1x & 9.008& 40.572& 124.753& 73.335
& 170.077& 60.443& 109.633& 64.461\%& 0.471
\\ 
   &3x & 8.981& 40.510& 125.153& 73.827
& 170.033& 60.734& 109.299& 64.281\%& 0.429
\\ 
   &5x & 8.957& 40.471& 125.489& 74.252
& 169.996& 60.981& 109.015& 64.128\%& 0.394
\\ 
   &7x & 8.937& 40.418& 125.782& 74.624
& 169.964& 61.191& 108.773& 63.998\%& 0.364
\\ 
   &9x & 8.919& 40.390& 126.048& 74.960
& 169.936& 61.380& 108.556& 63.881\%& 0.338
\\ 
\hline
\multirow{5}{*}{(c) $R_\text{ad}$}&1x & 9.008& 40.572& 124.753& 73.335
& 170.077& 60.443& 109.633& 64.461\%& 0.471
\\ 
   &3x & 8.853& 40.312& 127.211& 76.325
& 169.830& 62.087& 107.743& 63.442\%& 0.241
\\ 
   &5x & 8.799& 40.185& 128.061& 77.377
& 169.743& 62.660& 107.084& 63.086\%& 0.160
\\ 
   &7x & 8.771& 40.144& 128.472& 77.900
& 169.700& 62.948& 106.751& 62.906\%& 0.120
\\ 
   &9x & 8.755& 40.121& 128.720& 78.209
& 169.673& 63.126& 106.546& 62.795\%& 0.095
\\ 
\hline
\multirow{5}{*}{(d) $C_\text{a}$}&1x & 9.008& 40.572& 124.753& 73.335
& 170.077& 60.443& 109.633& 64.461\%& 0.471
\\ 
   &0.3x & 9.008& 40.572& 124.761& 73.343
& 170.076& 60.447& 109.630& 64.459\%& 0.470
\\ 
   &0.2x & 9.008& 40.563& 124.757& 73.334
& 170.076& 60.448& 109.628& 64.458\%& 0.470
\\ 
   &0.15x & 9.007& 40.567& 124.762& 73.336
& 170.076& 60.451& 109.625& 64.457\%& 0.470
\\ 
   &0.1x & 9.008& 40.566& 124.760& 73.338
& 170.076& 60.451& 109.625& 64.457\%& 0.470
\\ 
\hline
\multirow{5}{*}{(e) $C_\text{im}$}&1x & 9.008& 40.572& 124.753& 73.335
& 170.077& 60.443& 109.633& 64.461\%& 0.471
\\ 
   &0.3x & 9.012& 40.619& 124.736& 73.411
& 170.083& 60.401& 109.683& 64.488\%& 0.471
\\ 
   &0.2x & 9.014& 40.642& 124.693& 73.403
& 170.086& 60.379& 109.707& 64.501\%& 0.471
\\ 
   &0.15x & 9.015& 40.651& 124.665& 73.387
& 170.087& 60.372& 109.715& 64.505\%& 0.472
\\ 
   &0.1x & 9.015& 40.650& 124.640& 73.376
& 170.087& 60.372& 109.715& 64.505\%& 0.472
\\ 
\hline
\end{tabular}
}
\end{table}

\underline{Study 2: Perturbations across all resistances within a coronary LPM:} A healthy baseline condition was calibrated such that: cardiac output  $Q_{\text{mean}} = 4.981$ L/min, maximum cardiac output  $Q_{\text{max}} = 30.969$ L/min, systolic pressure $P_{\text{sys}} = 129.540$ mmHg, and diastolic pressure  $P_{\text{dia}} = 81.078$ mmHg.  The total left coronary flow in this state was 0.145 L/min (2.9 \% of the cardiac output). The values for all coronary LPM parameters of this healthy baseline condition are listed in Table \ref{tab:coronary_parameters_healthy}.

From this baseline state, all resistances ($R_{\text{a},i}$, $R_{\text{ap},i}$, $R_{\text{ad},i}$ in Fig. \ref{fig:CFD_model}) for each branch $i$ were simultaneously and proportionally increased to simulate higher degrees of CMD. Two scenarios were considered: a Moderate CMD case, produced by uniformly increasing $R_{\text{a},i}$, $R_{\text{ap},i}$ and $R_{\text{ad},i}$ by 100\% relative to the baseline; and a Severe CMD case, produced by uniformly increasing all resistances by 200\% relative to the baseline case.

\begin{figure}[h]
\centering
\includegraphics[width=0.8\textwidth]{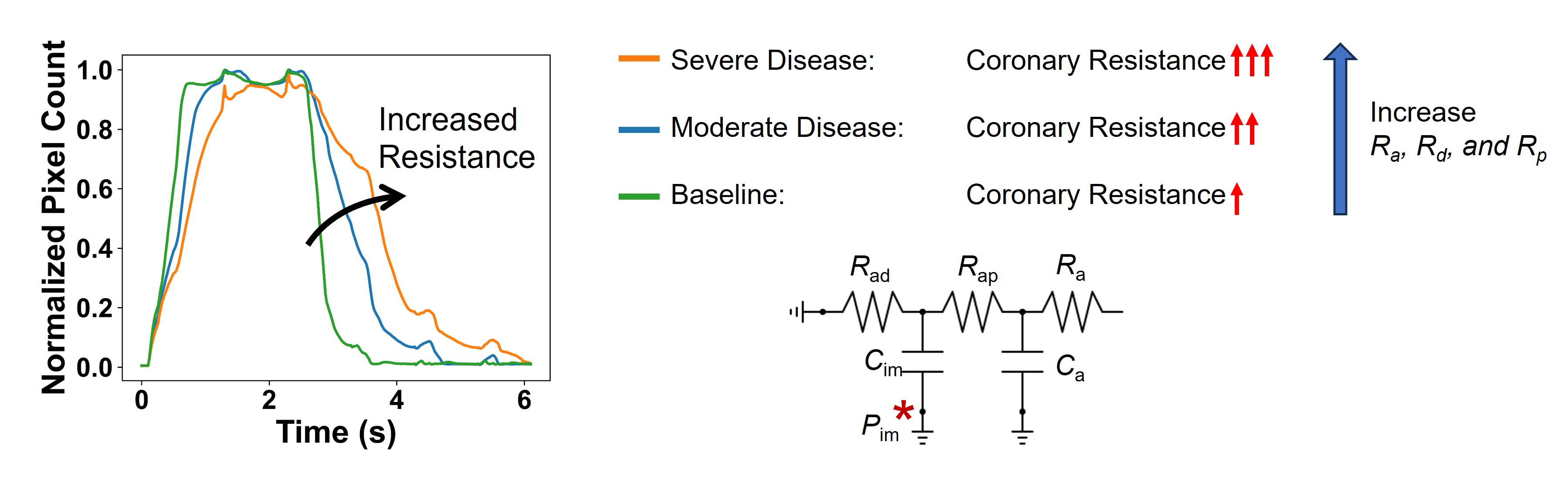}
\caption{\centering CIPs of simulations with increasing values of coronary resistance}
\label{fig:ComputationalCICComparison}
\end{figure}

Fig. \ref{fig:ComputationalCICComparison} shows CIPs for the baseline, moderate CMD, and severe CMD scenarios depicted in green, blue, and orange, respectively. Table \ref{tab:Hemo_numerical_CMD_analysis} lists corresponding hemodynamic metrics. Computational CIPs demonstrate significant variability in rising and falling slopes. Larger coronary microcirculation resistances are associated with slower rising and falling CIP slopes. The area under the CIP curves of the moderate and severe CMD cases are 14.571\% and 24.733\% larger than the baseline case, respectively. Additionally, larger coronary microcirculation resistance values lead to slight increases in systolic and diastolic pressures and slight reductions in cardiac output and ejection fraction. Furthermore, both moderate and severe CMDs result in a dramatic decrease in coronary flow, with reductions of 50\% and 66\%, respectively, compared to the baseline condition.

\begin{table}[H]
\centering
\footnotesize
\renewcommand{\arraystretch}{1.5} % Adjusting row spacing
\setlength{\tabcolsep}{5pt}
\caption{\centering Parameter values in coronary artery model for the healthy baseline case}
\label{tab:coronary_parameters_healthy}
\begin{tabular}{c c c c c c c}
\hline
  & $R_\text{a}$ & $R_\text{ap}$ & $R_\text{ad}$ & $C_\text{a}$ & $C_\text{im}$ \\ 
 & ($\text{Pa} \cdot \text{s} \cdot \text{mm}^{-3}$) & ($\text{Pa} \cdot \text{s} \cdot \text{mm}^{-3}$) & ($\text{Pa} \cdot \text{s} \cdot \text{mm}^{-3}$) & ($\text{mm}^3 \cdot \text{Pa}^{-1}$) & ($\text{mm}^3 \cdot \text{Pa}^{-1}$) \\
\hline
LAD &4.544& 1.363& 12.696& 0.014& 0.135
\\ 
OM1 &3.732& 1.120& 10.429& 0.014& 0.135
\\ 
OM2 &7.153& 2.146& 19.989& 0.007& 0.135
\\
LCx &6.398& 1.919& 17.878& 0.007& 0.135
\\
AM &4.757& 1.427& 13.293& 0.012& 0.189
\\
RCA &3.199& 0.960& 8.939& 0.012& 0.189
\\ 
     
\hline
\end{tabular}

\end{table}

\begin{table}[H]
\centering
\footnotesize
\renewcommand{\arraystretch}{1.5} % Adjusting row spacing
\setlength{\tabcolsep}{5pt}
\caption{\centering Hemodynamic metrics of simulations with increasing values of coronary resistance}
\label{tab:Hemo_numerical_CMD_analysis}
\resizebox{\textwidth}{!}{  % Automatically resize the table to fit within the text width
\begin{tabular}{c c c c c c c c c c}
\hline
 & $Q_{\text{mean}}$ & $Q_{\text{max}}$ & $P_{\text{sys}}$  & $P_{\text{dia}}$ & $\text{EDV}$& $\text{ESV}$& SV & EF& $Q_\text{left tree}$\\ 
 & (L/min) & (L/min) & (mmHg) & (mmHg) & (ml) & (ml)  & (ml) &  &(L/min)\\
\hline
Baseline & 4.981& 30.969& 129.540& 81.078
& 160.607& 77.567& 83.041& 51.704\%& 0.145
\\ 
Moderate CMD & 4.925& 30.766& 130.524& 82.372
& 160.479& 78.385& 82.095& 51.156\%& 0.075
\\ 
Servere CMD & 4.906& 30.698& 130.876& 82.835
& 160.436& 78.662& 81.775& 50.970\%& 0.050
\\ 
\hline
\end{tabular}
}
\end{table}

\section{Discussion}\label{Discussion}

\underline{Model calibration for rest and hyperemia:} Model calibration was performed for rest and hyperemia using a combination of patient-specific and literature data. The impact of adenosine on cardiovascular function is complex \cite{guieu2020adenosine}, generally leading to reductions in heart rate. However, for the dose typically used for inducing hyperemia in the cath lab (140 $\mu$g/kg/min), an increase in heart rate is observed \cite{rongen1999effect}, in addition to coronary and systemic vasodilation. 

Systemic vasodilation leads to increases in vessel diameter which corresponds with decreased proximal systemic resistance $R_\text{sp}$, and distal systemic resistance $R_\text{sd}$ of the aortic LPM. Furthermore, vasodilation stretches the vessel wall, decreasing systemic compliance $C_\text{s}$. Adenosine-induced vasodilation therefore leads to reductions in blood pressure, which trigger an increase in sympathetic tone, subsequently raising heart rate. Adenosine-induced hyperemia increases maximum elastance $E_\text{max}$, time to maximum elastance $t_\text{max}$, and left atrial pressure $P_\text{LA}$ \cite{casas2018non}, leading to enhanced heart contractility and increased cardiac output. Since minimum elastance $E_\text{min}$ reflects the passive mechanical properties of the ventricle, we assumed it is constant between resting and hyperemic states. The combination of increased left atrial pressure $P_\text{LA}$ and constant minimum elastance $E_\text{min}$ results in an increase in $\text{EDV}$. Additionally, the rise in maximum elastance ($E_\text{max}$) causes a decrease in $\text{ESV}$. Increased $\text{EDV}$ and decreased $\text{ESV}$ lead to increased SV and EF. The shortening of the cardiac cycle occurs primarily during diastole. Consequently, these changes in the heart and aorta LPMs result in an increased cardiac output $Q_{\text{mean}}$, driven by a rise in SV and heart rate. Furthermore, reduced afterload causes both systolic $P_\text{sys}$ and diastolic aortic pressures $P_\text{dia}$ to decrease.

Coronary vasodilation during hyperemia results in a decrease of both resistance and capacitance, and a corresponding increase in flow. Specifically, the LAD flow increased by a factor of 2.203, consistent with patient data. During model calibration, we assumed that all coronary LPMs exhibited a similar degree of microvascular disease. Therefore, based on Murray’s Law, flow rate $Q_i$ at branch $i$ was assumed proportional to $r_i^3$. While these assumptions simplify the calibration process, they may not hold true in situations of localized microvascular disease with increased resistance in a subset of the coronary branches. These assumptions may contribute to the discrepancies observed between clinical angiograms and computational predictions in Figs. \ref{fig:Angios_comparison_resting} and \ref{fig:Angios_comparison_hyper}. 

The calibrated computational CIPs accurately capture the general trends of the patient-specific clinical CIPs but exhibit some discrepancies around the plateau of the profile, both for rest and hyperemia (Fig. \ref{fig:CIP_comparison}). These discrepancies reflect the differences between computational and clinical angiograms over time (see Figs. \ref{fig:Angios_comparison_resting} and \ref{fig:Angios_comparison_hyper}). There are several factors contributing to these discrepancies: (1) The AngioNet segmentation algorithm \cite{iyer2021angionet} is not perfect. Vessels with moderate contrast concentration may be included in the segmentation in one frame but excluded in the next, leading to the oscillations seen in the clinical CIP. (2) Vessel motion is absent in the computational angiograms, but significant in the clinical angiograms. Vessel motion results in rotation and translation in and out of the plane of the X-ray camera, which is not accounted for by the computational model. (3) The 3D geometry of the computational model does not include all the branches of the coronary tree (e.g., see the comparison between branches in clinical angiogram vs. computational angiogram at $t=1.4$ s for Fig. \ref{fig:Angios_comparison_resting}). (4) Lastly, 
as previously mentioned, the assumption of uniform microvascular health across all branches is not generally true, and therefore the speed of contrast washout may differ significantly across different branches. 

\underline{Sensitivity studies:} Studies were performed by perturbing both individual coronary LPMs parameters, as well as all resistances within a given coronary LPM. The individual parameter sensitivity studies were performed around a typical hyperemic hemodynamic condition and revealed that resistance, compared to capacitance, has a more substantial impact on the rising and falling slopes of CIP. Resistance greatly determines the flow rate, and therefore the advective component of the transport of the contrast agent. In contrast, capacitance primarily modulates the shape of the flow waveform and has minimal influence on the flow rate, and therefore the CIP slopes. Moreover, the parameter $R_\text{ad}$ representing the distal small artery resistance was found to have a pronounced effect on the CIP shape, particularly through its role in backflow, with higher $R_\text{ad}$ values producing a prominent bump in the CIP around $t\sim 2$ s. These results highlight the nuanced contributions of resistance and capacitance in CIP formation and underscore the potential of resistance, particularly the most distal small arterial resistance, as diagnostic markers for microvascular dysfunction.

The sensitivity studies which uniformly altered all resistances within each coronary LPM were performed around a typical baseline hemodynamic condition. Results revealed that changes in resistances at all depths of the coronary microcirculation ($R_a$, $R_\text{ap}$, $R_\text{ad}$) have a substantial impact on the shape of the CIPs. The slower rising and falling slopes associated with increased resistances reflect a higher degree of CMD. The patterns in Fig. \ref{fig:ComputationalCICComparison} are consistent with those presented in the clinical CIPs depicted in Fig. \ref{fig:Jesse_paper}. The distinct patterns of CIP dynamics observed for the healthy, moderate, and severe disease scenarios also suggest that computational simulations of contrast injection can serve as a valuable tool for analyzing CMD.

\section{Conclusion and Future Work}\label{Conclusion}

In this paper, we have presented a multi-physics model of contrast injection to study the relationship between features of contrast washout in coronary angiography videos and microvascular health. Specifically, we are interested in studying the relationship between microvascular parameters (resistance, compliance) and the contrast intensity profile (CIP) \cite{resnick2024neural}, which quantifies the amount of contrast agent in the epicardial vessels on a frame-by-frame basis through segmentation of the angiography data.   The multi-physics model features 3D Navier-Stokes and scalar advection-diffusion equations, combined with lumped parameter models (LPM) describing the function of left heart, distal aorta, and coronary circulations. We have demonstrated that: 1) The multi-physics model can be effectively calibrated to match arbitrary sets of clinical data consisting of coronary angiograms and data on flow and pressure, under resting and hyperemic conditions. The model produces computational CIPs which closely mimic CIPs derived from clinical angiograms (see Fig. \ref{fig:Jesse_paper}); 2) Coronary microvascular resistances of the multi-physics model determine the slopes of the CIPs.  

Model calibration was performed through a two-stage optimization process. The first stage involved tuning the heart and aorta LPMs to match patient-specific hemodynamic data, using a 'pure 0D' model with minimal computational cost, and a differential evolution optimization scheme. In the second stage, coronary artery LPMs were adjusted to achieve target flow rates and waveforms, initially through 0-D simulations. Then, grid searches were conducted to fine-tune the coronary artery LPMs through multi-physics simulations, ensuring agreement between computational and clinical CIPs. Following the calibration process, sensitivity studies were performed to explore the impact of coronary LPM parameters on the shape of the CIP, and to investigate the relationship between CIPs and the degree of CMD. The results demonstrate that the calibrated multi-physics simulations produce physiologically relevant hemodynamic results, such as flow and pressure waveforms, while having the computational angiograms and CIPs closely match their clinical counterparts. These findings demonstrate that the proposed model is a reliable surrogate for simulating contrast injection and washout in clinical angiography. Sensitivity studies further revealed that resistance has a more significant effect on the rising and falling slopes of CIP than capacitance. Slower rising and falling slopes in the CIP signify higher resistances in the microcirculation, indicating greater severity of CMD.

This work has several limitations. During the model calibration stage, only a limited set of patient-specific data was utilized, including angiograms, their corresponding CIPs ($\hat{\text{CIP}}$) and invasive CFR measurement ($\hat{\text{CFR}}$). However, certain target flow and pressure information, such as mean and peak cardiac outputs ($\hat{Q}_{\text{mean}}$ and $\hat{Q}_{\text{max}}$), systolic and diastolic blood pressure ($\hat{P}_{\text{sys}}$ and $\hat{P}_{\text{dia}}$), as well as lower bound and upper bound left ventricular end-diastolic volume ($\hat{\text{EDV}}_{\text{LB}}$ and $\hat{\text{EDV}}_{\text{UB}}$), were extracted from literature. This mismatch in data sources led to inconsistencies and ultimately resulted in inaccurate simulation results. Furthermore, due to a lack of data on individual flows for each coronary vessel, Murray's law was assumed to determine flow splits, which is one of the targets for the coronary artery model calibration. However, this assumption may not be valid for specific patients, leading to discrepancies in contrast injection and washout at vessel branches. A possible method to remedy this shortcoming is to extract bulk flow (mean flow over the cardiac cycle) for each individual branch from the coronary videos. This is an active area of research in our group. Additionally, the 3D-0D coupled multi-physics model does not account for heart motion effects on the coronary arteries, and simulates the vessel walls as rigid. These simplifications contribute to discrepancies between computational and clinical data, such as angiograms, segmentation masks, and CIPs. However, most of the coronary flow occurs in diastole, when the heart motion is minimal. Furthermore, simulation of coronary hemodynamics using fixed grids and rigid walls is quite common and has been used successfully to predict clinically relevant quantities (e.g., FFR-CT \cite{taylor2013computational}). Fluid-structure interaction models, accounting for large motion of the coronary vessels, could be used to improve the fidelity of the proposed model. Lastly, the relationship between resistance and compliance in coronary circulation is far more complex than  has been discussed here. As CMD progresses, vessel rarefaction results in simultaneous increases in microvascular resistance and decreases in microvascular compliance \cite{mohammed2015coronary}, which would be region-specific. Therefore, resistances and compliances should not be modeled as independent variables to simulate CMD progression. 

We submit that the model presented here and these results are potentially transformative, as they provide a tool for interpreting angiographic data and ultimately extracting information concerning coronary microcirculation. Furthermore, this paper is, to our knowledge, the first effort to simulate and analyze the complex physics of coronary angiography with a high-fidelity multiphysics computational model. Future research will focus on developing data-driven machine learning models to analyze features in CIPs and use them to predict invasive measurements, such as coronary flow reserve (CFR) and the index of microcirculatory resistance (IMR).

\section*{Acknowledgments}
This study received no funding.
\section*{Declarations}

\textbf{Competing interests}:

Authors CAF and BKN are founders and shareholders of AngioInsight, Inc. All other authors declare no financial or non-financial competing interests. 

\textbf{Data availability}:

The datasets analyzed during the current study are not publicly available because they are patient medical images, such as coronary CTA and angiograms.

\textbf{Code availability}:

The code for multiphysics simulation of contrast injection is available in 
www.crimson.software.

Source code can be accessed via https://github.com/carthurs/CRIMSONFlowsolver (CRIMSON Flowsolver) under the GPL v3.0 license, and https://github.com/carthurs/CRIMSONGUI (CRIMSON GUI).

The underlying code for model calibration is not publicly available but may be made available to qualified researchers on reasonable request from the corresponding author.

\textbf{Author contributions}:

H.Y., B.K.N., K.G., and C.A.F. conceived and designed the study. H.Y., J.Z., and I.Z.A. developed the multi-physics model and conducted simulations. H.Y. and J.Z. carried out model calibration and sensitivity analysis. H.Y., K.G., and C.A.F. drafted the manuscript, with all authors contributing to its review and approval.

\bibliography{sample.bib}
\end{document}